\RequirePackage{arydshln}
\documentclass[aps,nofootinbib,superscriptaddress,preprintnumbers,pra,10pt,notitlepage]{revtex4-1}

\usepackage{colortbl} 
\definecolor{Gray}{gray}{0.95}
\definecolor{RGray}{gray}{0.93}
\definecolor{CGray}{gray}{0.92}

\usepackage{amsmath,amssymb,amsfonts, bm,bbm,slashed}
\usepackage{dsfont} 
\usepackage[hidelinks]{hyperref}
\usepackage{graphicx}
\usepackage{enumitem}
\usepackage{arydshln}
\usepackage{mathtools}
\usepackage{bbold}
\usepackage{adjustbox}
\usepackage{multirow}
\usepackage{tikz}
\usetikzlibrary{calc,tikzmark,fit,shapes.geometric,matrix,decorations.markings,arrows.meta,decorations.pathmorphing,patterns,positioning,snakes}
\usepackage{soul}
\usepackage[capitalize]{cleveref}
\usepackage{gensymb}

\makeatletter
\g@addto@macro\bfseries{\boldmath}
\makeatother

\newcommand{\be} {\begin{equation}}
\newcommand{\ee} {\end{equation}}
\newcommand{\bea} {\begin{eqnarray}}
\newcommand{\eea} {\end{eqnarray}}

\newcommand{\ba} {\begin{array}}
\newcommand{\ea} {\end{array}}
\newcommand{\gsim}{\lower.7ex\hbox{$\;\stackrel{\textstyle>}{\sim}\;$}}
\newcommand{\lsim}{\lower.7ex\hbox{$\;\stackrel{\textstyle<}{\sim}\;$}}

\newcommand{\cC}{\mathcal{C}}
\newcommand{\cO}{\mathcal{O}}

\newcommand{\cB}{\mathcal{B}}
\newcommand{\cL}{\mathcal{L}}
\newcommand{\xZ}{x_{\Zp}}
\newcommand{\xG}{x_{G^\prime}}
\newcommand{\mU}{m_U}
\newcommand{\LU}{\Lambda_U}
\newcommand{\mZ}{m_{Z^\prime}}
\newcommand{\mG}{m_{G^\prime}}
\newcommand{\Gp}{G^\prime}
\newcommand{\Zp}{Z^\prime}

\tikzset
  {midarrow/.style={decoration={markings,mark=at position 0.5 with
     {\arrow[thin,xshift=2pt]{Triangle[length=4pt,#1]}}},postaction={decorate}}
  }

\tikzset{
proton/.style = {circle, draw=black, thin, fill=black!20!white, minimum size=#1,
              inner sep=0pt, outer sep=0pt},
proton/.default = 6pt 
}

\tikzset{
blob/.style = {circle, draw=black, thin, preaction={fill, black!20!white}, pattern=north east lines, minimum size=#1,
              inner sep=0pt, outer sep=0pt},
blob/.default = 6pt 
}

\tikzset{
wc/.style = {circle, fill, minimum size=#1,
              inner sep=0pt, outer sep=0pt},
wc/.default = 4pt 
}

\tikzset{vector/.style={decorate, decoration=snake}}

\begin{document}

\preprint{ZU-TH 11/23}

\title{Third-Family Quark–Lepton Unification and Electroweak Precision Tests}
 
\author{Lukas Allwicher}
\email{lukall@physik.uzh.ch}
\affiliation{Physik-Institut, Universit\"at Z\"urich, CH-8057 Z\"urich, Switzerland}
\author{Gino Isidori}
\email{isidori@physik.uzh.ch}
\affiliation{Physik-Institut, Universit\"at Z\"urich, CH-8057 Z\"urich, Switzerland}
\author{Javier M. Lizana}
\email{jlizana@physik.uzh.ch}
\affiliation{Physik-Institut, Universit\"at Z\"urich, CH-8057 Z\"urich, Switzerland}
\author{Nud{\v z}eim Selimovi{\'c}}
\email{nudzeim@physik.uzh.ch}
\affiliation{Physik-Institut, Universit\"at Z\"urich, CH-8057 Z\"urich, Switzerland}
\author{Ben A. Stefanek}
\email{ben.stefanek@physik.uzh.ch}
\affiliation{Physik-Institut, Universit\"at Z\"urich, CH-8057 Z\"urich, Switzerland}

\begin{abstract}
\vspace{5mm}
We analyze the compatibility of the hypothesis of third-family quark-lepton unification at the TeV scale with electroweak precision data, lepton flavor universality tests, and high-$p_T$ constraints. We work within the framework of the UV complete flavor non-universal 4321 gauge model, which is matched at one loop to the Standard Model Effective Field Theory. For consistency, all electroweak precision observables are also computed at one loop within the effective field theory. At tree level, the most sizeable corrections are to $W\rightarrow \tau\nu_\tau$ and 
$Z \to \nu_\tau \nu_\tau$
due to integrating out a pseudo-Dirac singlet fermion required by the model for neutrino mass generation. At loop level,
the new colored states 
of the model generate large flavor-universal contributions to the electroweak precision observables via leading- and next-to-leading log running effects, yielding a significant improvement in the electroweak fit (including an increase in the $W$-boson mass). These effects cannot be decoupled if the model addresses the charged-current $B$-meson anomalies.  Overall, we find good compatibility between the data sets, while simultaneously satisfying all low- and high-energy constraints.

\end{abstract}

\maketitle
\section{Introduction}\label{sec:intro}

The large amount of data constraining physics beyond the Standard Model (SM) has significantly increased the pressure on models addressing the Higgs hierarchy problem via TeV-scale new physics (NP). However, a closer inspection of present high-energy bounds shows that they are quite different depending on the flavor structure of the hypothetical new states: NP coupled universally to the SM fermions (or only to the light families) is strongly constrained by LHC searches, with mass bounds typically in the several TeV range.
On the contrary, the bounds on new states coupled dominantly to the third generation are significantly weaker, barely exceeding 1~TeV in several well-motivated cases.
Given that the light families are weakly coupled to the Higgs sector, 
the possibility of addressing the Higgs hierarchy problem via new states with flavor non-universal interactions lying just above the TeV scale remains an attractive possibility.

An independent but complementary indication toward new flavor non-universal interactions comes from the flavor sector itself. Also in this case the strongest NP bounds are those derived from the light fermion families. Moreover, the natural assumption that the highly hierarchical SM Yukawa couplings are the result of some beyond-the-SM interactions points toward the existence of flavor non-universality in the ultraviolet (UV) completion of the SM.

An interesting way to combine these two general indications is realized by postulating new flavor non-universal gauge interactions at the TeV scale, with the non-universality distinguishing the third generation. This idea has received a lot of interest in the last few years in connection with a series of hints of deviations from the SM in $B$-meson decays~\cite{Bordone:2017bld,Greljo:2018tuh,Fuentes-Martin:2020pww,Fuentes-Martin:2020bnh,Fuentes-Martin:2022xnb,Davighi:2022bqf}. However, it is worth stressing that the appeal goes beyond the $B$-meson anomalies and connects to the older, more general idea of a (flavor-based) multi-scale structure in the UV completion of the SM~\cite{Dvali:2000ha,Panico:2016ull,Allwicher:2020esa,Barbieri:2021wrc}.

The common aspect of all the recent flavor non-universal proposals in Ref.~\cite{Bordone:2017bld,Greljo:2018tuh,Fuentes-Martin:2020pww,Fuentes-Martin:2020bnh,Fuentes-Martin:2022xnb,Davighi:2022bqf} is the hypothesis that TeV-scale dynamics is ruled by the extended gauge group $SU(4)_h \times SU(3)_l \times SU(2)_L \times U(1)_X$. Commonly denoted the ``4321" group, it was first proposed in a flavor universal version in~\cite{DiLuzio:2017vat} and then later in a  non-universal version featuring  quark-lepton unification \`a la Pati-Salam~\cite{Pati:1974yy} for the third family SM fermions~\cite{Bordone:2017bld,Greljo:2018tuh}. Besides the clear theoretical appeal corresponding to an explanation of the observed flavor hierarchies~\cite{Bordone:2017bld},
quark-lepton unification~\cite{Greljo:2018tuh}, and the possibility to address the 
Higgs hierarchy problem~\cite{Fuentes-Martin:2020bnh,Fuentes-Martin:2022xnb}, the key phenomenological property of the 4321 setup is the presence of a TeV-scale $U_1$ gauge vector leptoquark (LQ). Before any of these model-building attempts, this particle was recognized as the most efficient mediator to explain all the anomalies in semi-leptonic $B$ decays~\cite{Alonso:2015sja,Calibbi:2015kma,Barbieri:2015yvd,Bhattacharya:2016mcc,Buttazzo:2017ixm}. Its main phenomenological virtue is addressing the charged-current
$b\to c\ell \nu$ anomaly~\cite{BaBar:2013mob,Belle:2016dyj,LHCb:2015gmp,LHCb:2017smo,LHCb:2017rln} 
that defines the scale of the model while being consistent with all available low- and high-energy constraints. In this respect, the updated analysis of lepton flavor universality (LFU) violations in the $b\to s\bar \ell \ell$~\cite{LHCb:2022qnv} system changes little 
concerning the theoretical motivation of and interest in the model, although there is clearly less experimental evidence for a large NP effect in light-family leptons. 

The rich phenomenology following from the quark-lepton unification hypothesis has been studied in detail in several previous works, including implications for $\tau$ decays~\cite{Allwicher:2021ndi}, $\Delta F=2$ and $\Delta F=1$ transitions~\cite{DiLuzio:2018zxy,Fuentes-Martin:2020hvc,Crosas:2022quq}, 
and high-$p_T$ collider signatures~\cite{Allwicher:2022gkm,Baker:2019sli,Buonocore:2020erb,Buonocore:2022msy}.
An element that has been missing so far, however, is a systematic study of the implications for electroweak precision observables (EWPO),
which constitutes the main objective of this paper.

The interplay between semi-leptonic interactions at the TeV scale,
generated by flavor non-universal interactions, and EWPO has already 
been pointed out in a pure Effective Field Theory (EFT) context~\cite{Feruglio:2016gvd,
Feruglio:2017rjo}: the running of the (non-universal) four-fermion operators into Higgs-fermion operators
results in sizable corrections to the EWPO.
Having a UV-complete model at hand, we now go beyond these initial studies including all the relevant finite pieces resulting from the matching of the 4321 
model into the EFT of the SM (the so-called SMEFT). 
In this respect, we analyze for the first time in a systematic way 
the effects of new  fermion fields (and related interactions), which  
are present in all realistic scenarios featuring 
third-family quark-lepton unification at the TeV scale.
First, these models necessarily include vector-like fermions 
in order to describe the mixing between the light and heavy SM families. Second, an inevitable consequence of unifying quarks and leptons of the third family is the prediction of a degenerate top quark and tau neutrino.  In order to resolve this issue, an inverse-seesaw mechanism with additional gauge singlet fermions is a necessary ingredient~\cite{Greljo:2018tuh}. In addition to the $U_1$ LQ, the model requires the existence of two massive neutral vector bosons. As we shall show, all of these new states generate important effects in EWPO. For example, the exotic fermions in many cases induce one-loop matching contributions that can compete in size with pure renormalization group (RG) effects. Additionally, four-quark operators generated after integrating out the neutral vectors can run beyond the leading-log approximation into relevant contributions for the electroweak (EW) fit.
 Besides providing analytical expressions to evaluate all these effects in general terms, we perform a detailed phenomenological analysis of 
different EWPO, identifying the most interesting ones to further test this hypothesis in the future. Finally, we analyze the compatibility of EWPO with LFU tests in $b\rightarrow c\tau\nu$ and $\tau$-decays, as well as with high-$p_T$ constraints by performing a global fit to the data.

The paper is organized as follows. In Section~\ref{sec:model} we define the non-universal 4321 model.
In Section~\ref{sec:nlo} we discuss the matching procedure from the non-universal 4321 model to the SMEFT, the main running effects from the 4321 breaking scale down to the electroweak scale, and the one-loop computation of the EWPO in the EFT. The phenomenological discussion as well as the results of the global fit are presented in 
Section~\ref{sec:pheno} and summarized in the Conclusions. A series of appendices contain more technical aspects: a detailed discussion about one-loop matching contributions (\ref{app:AppendixA}); a discussion and analytical expressions for the RG running of the Wilson coefficients (WCs) (\ref{app:AppendixB}); one-loop expressions for the EWPO in terms of SMEFT coefficients ({\ref{app:EWPO}); phenomenological expressions for the LFU ratios in $b\to c\ell \nu$ transitions (\ref{app:lowenergy}).

\section{The model}\label{sec:model}
As already anticipated, the gauge group of the model is 
\begin{equation}
\mathcal{G}_{4321} = SU(4)_h \times SU(3)_l \times SU(2)_L \times U(1)_X\,,
\end{equation}
where $SU(3)_c$ is identified as the diagonal subgroup of $SU(4)_h \times SU(3)_l$ and the flavor-universal group $SU(2)_L$ acts like in the SM.
The hypercharge is given by
$Y=X+\sqrt{2/3}\,T_4^{15}$, where $T_4^{15}=\frac{1}{2\sqrt{6}}\mathrm{diag}(1,1,1,-3)$ is a generator
of $SU(4)_h$, meaning that $U(1)_X$ coincides with hypercharge for the light families (see~\cref{tab:fieldcontent} for the full matter field content). 
Without mass mixing, the three  $SU(4)_h$ multiplets $\psi_L$ and $\psi_R^\pm$ can be identified with the 
SM third-generation fermions (with the addition of a right-handed neutrino). 
The only exotic fermions are a
vector-like fermion $\chi_{L,R} = (Q_{L,R},L_{L,R})$, with the same gauge quantum numbers as $\psi_L$, and a gauge singlet $S_L$ needed to achieve an acceptable tau neutrino mass.

\begin{table}[t]
\begin{center}
\renewcommand{\arraystretch}{1.2}
\begin{tabular}{|c|c|c|c|c|}
\hline
Field & $SU(4)_h$ & $SU(3)_l$ & $SU(2)_L$ & $U(1)_{X}$ \\
\hline
\hline
$\psi_L$ & $\mathbf{4}$ & $\mathbf{1}$ & $\mathbf{2}$ & $0$ \\ 
$\psi_R^{\pm}$ & $\mathbf{4}$ & $\mathbf{1}$ & $\mathbf{1}$ & $\pm1/2$ \\
$\chi_{L,R}$ & $\mathbf{4}$ & $\mathbf{1}$ & $\mathbf{2}$ & 0 \\ \hline
$q_L^{i}$ & $\mathbf{1}$ & $\mathbf{3}$ & $\mathbf{2}$ & $1/6$ \\
$u_R^{i}$ & $\mathbf{1}$ & $\mathbf{3}$ & $\mathbf{1}$ & $2/3$ \\
$d_R^{i}$ & $\mathbf{1}$ & $\mathbf{3}$ & $\mathbf{1}$ & $-1/3$ \\
$\ell_L^{i}$ & $\mathbf{1}$ & $\mathbf{1}$ & $\mathbf{2}$ & $-1/2$ \\
$e_R^{i}$ & $\mathbf{1}$ & $\mathbf{1}$ & $\mathbf{1}$ & $-1$ \\ \hline
$S_L$ & $\mathbf{1}$ & $\mathbf{1}$ & $\mathbf{1}$ & $0$ \\ \hline
$\Omega_1$ & $\bar{\mathbf{4}}$ & $\mathbf{1}$ & $\mathbf{1}$ & $-1/2$ \\
$\Omega_3$ & $\bar{\mathbf{4}}$ & $\mathbf{3}$ & $\mathbf{1}$ & $1/6$ \\
$\Omega_{15}$ & $\mathbf{15}$ & $\mathbf{1}$ & $\mathbf{1}$ & $0$ \\
$H$ & $\mathbf{1}$ & $\mathbf{1}$ & $\mathbf{2}$ & $1/2$ \\ \hline
\end{tabular}
\end{center}
\caption{Field content of the model. The would-be third generation quarks and leptons are unified in $\psi_L\equiv (q_L^{\prime 3}\,\, \ell_L^{\prime 3})^\intercal$, $\psi_R^+\equiv(u_R^3\,\, \nu_R^3)^\intercal$, and $\psi_R^-\equiv(d_R^3\,\, e_R^3)^\intercal$, while $i=1,2$ for the $SU(4)_h$ singlets.
}
\label{tab:fieldcontent}
\end{table}

The scalar fields $\Omega_1$, $\Omega_3$, and $\Omega_{15}$ mediate the $\mathcal{G}_{4321}\to$~SM breaking at the TeV scale.\footnote{In this version of the model, the scalar field $\Omega_{15}$ is necessary to induce a mixing between the $SU(4)_h$-charged fields. In alternative versions of the model, this can also be achieved by charging differently the vector-like fermion $\chi_{L,R}$~\cite{Fuentes-Martin:2020bnh}.} This results in three heavy gauge fields,   
transforming under the SM as $U_1\sim(\bf{3},\bf{1},2/3)$, $G^\prime\sim(\bf{8},\bf{1},0)$ and $Z^\prime\sim(\bf{1},\bf{1},0)$.  
More details about symmetry breaking can be found in~\cite{Fuentes-Martin:2020hvc}.
In the $\mathcal{G}_{4321}$ broken phase, a mixing between chiral and vector-like fermions is generated. Defining $\Psi_{q}^{\prime \intercal} = (q_{L}^{\prime 1} \;q_{L}^{\prime 2} \; q_L^{\prime 3} \;  Q^{\prime}_L)$ and $\Psi_\ell^{\prime \intercal} = (\ell_{L}^{\prime 1} \; \ell_{L}^{\prime 2} \; \ell_L^{\prime 3} \;  L^{\prime}_L)$, the mass mixing can be written as
\begin{equation}
-\mathcal{L} \supset \bar \Psi_{q}^{\prime} \, \mathbf{M}_{q}  Q_R + \bar \Psi_{\ell}^{\prime} \, \mathbf{M}_{\ell}  L_R + \textrm{h.c.} \,,
\end{equation}
where, without loss of generality, the mass vector can be decomposed as
\begin{align}
\mathbf{M}_{q,\ell} &= \mathbf{W}_{q,\ell}^{\dagger} \, {\bf O}_{q,\ell}^{\dagger} \,\, (0\;\, 0\;\, 0\;\, m_{Q,L})^{\intercal}
= \begin{pmatrix}
\mathbb{1}_{2\times 2} & 0_{2\times 2}  \\
0_{2\times 2} & W_{q,\ell}^\dagger  \\
\end{pmatrix}\begin{pmatrix}
1 & 0  & 0 & 0\\
0 & c_{q,\ell}  & 0 & s_{q,\ell}\\
0 & 0 & 1 & 0  \\
0 & -s_{q,\ell} & 0 & c_{q,\ell}
\end{pmatrix}\begin{pmatrix}
0\\
0\\
0\\
m_{Q,L}
\end{pmatrix}\,,
\end{align}
with $m_Q$ and $m_L$ being the physical vector-like fermion masses. Here, ${\bf O}_{q,\ell}$
describe the mixing between different $SU(4)_h$ representations ($q_L^{\prime 2}\,-\,Q_L^\prime$ and $\ell_L^{\prime 2}\,-\,L_L^\prime$), parameterized by the angle $\theta_{q,\ell}$, while ${\bf W}_{q,\ell}$
parameterizes the mixing amongst $SU(4)_h$ states ($q_L^{\prime 3}\,-\,Q_L^\prime$ and $\ell_L^{\prime 3}\,-\,L_L^\prime$). After moving to the mass basis, $\Psi_{q,\ell}^{\prime} \rightarrow \mathbf{W}_{q,\ell}^{\dagger} \, {\bf O}_{q,\ell}^{\dagger}  \Psi_{q,\ell}$, the relevant interactions of the fermions with the massive gauge bosons read
\begin{align}
\label{eq:ULag}
\mathcal{L}_{U_1,G^{\prime},Z^\prime}  &\supset\frac{g_4}{\sqrt{2}} U_1^\mu\left[\, \beta_L  \, \bar \Psi_q  \gamma_\mu \Psi_{\ell} + e^{i\phi_R} \, \bar{b}_R \gamma_\mu \tau_R\right] + g_4 G^{\prime\mu} \left[\kappa_q \,\bar \Psi_q \gamma_\mu\Psi_q + \bar u_R^3 \gamma_\mu u_R^3 + \bar b_R^3 \gamma_\mu b_R^3\right]\nonumber\\
& + \frac{g_4}{2\sqrt{6}} Z^{\prime\mu} \left[\,\kappa_q\,\bar \Psi_q  \gamma_\mu \Psi_q - 3\, \xi_\ell\, \bar \Psi_\ell \gamma_\mu \Psi_\ell + \bar u_R^3 \gamma_\mu u_R^3 + \bar b_R^3 \gamma_\mu b_R^3 -  3\,\bar \tau_R^3 \gamma_\mu \tau_R^3\right]\,,
\end{align}
where $g_4$ indicates the $SU(4)$ gauge coupling, $\phi_R$ is the relative phase between left- and right-handed $U_1$ currents, and $G^{\prime}_\mu = G^{\prime a}_\mu T^a$ with $T^a$ being the Gell-Mann matrices. The flavor structure is summarised in the matrices
\begin{align}
\kappa_q &= {\bf O}_{q} \kappa_q^\prime {\bf O}_{q}^{\dagger} \,, \hspace{8mm} \kappa_q^\prime=\mathrm{diag}(0,0,1,1)\,, \\
\xi_\ell &= {\bf O}_{\ell} \, \xi_{\ell}^\prime \, {\bf O}_{\ell}^{\dagger} \,, \, \hspace{8mm} \xi_\ell^\prime=\mathrm{diag}(0,0,1,1)\,,
\end{align}
\begin{align}
\label{eq:betaL&W}
\beta_L = {\bf O}_{q} {\bf W} \beta_L^\prime {\bf O}_{\ell}^{\dagger} \,, \hspace{8mm} 
{\bf W} = 
\begin{pmatrix}
\mathbb{1}_{2\times 2} & 0_{2\times 2}  \\
0_{2\times 2} & W  \\
\end{pmatrix} \,,
\hspace{8mm}
\beta_L^\prime=\mathrm{diag}(0,0,1,1)
\,,
\end{align}
where we have neglected terms $O(g_s^2/g_4^2)$ and $O(g_Y^2/g_4^2)$, where $g_s$ and $g_Y$ are the SM strong and hypercharge couplings.
By re-phasing $q_L^3$, $\ell_L^3$, and the left-handed vector-like fermion fields, the matrix $W= W_q  W_\ell^\dagger$ can be written as a $2\times 2$ real rotation, parameterized by an angle $\chi$,
\begin{equation}\label{eq:Wmatrix}
W = \begin{pmatrix}
c_{\chi} & s_{\chi}  \\
-s_{\chi} & c_{\chi}  \\
\end{pmatrix}   \,.
\end{equation}

\noindent Similarly, we can write the Yukawa interactions in the mass basis as
\begin{align}
-\mathcal{L}_Y  &\supset \bar \Psi_q  {\bf Y}_u \tilde{H} u_R^3 + \bar \Psi_q  {\bf Y}_d H d_R^3 +\bar \Psi_q  {\bf Y}_\nu \tilde{H} \nu_R^3 + \bar \Psi_q  {\bf Y}_e H e_R^3 +\rm{h.c.}\,,
\end{align} 
where the Yukawa vectors can be expressed as
\begin{align}
    {\bf Y}_{u,d} = {\bf O}_{q} \begin{pmatrix}
    0\\
    0\\
    y_{t,b}\vspace{1mm}\\
    Y_\pm \end{pmatrix}\,,\quad {\bf Y}_{\nu,e} = {\bf O}_{\ell}{\bf W}^\dagger \begin{pmatrix}
    0\\
    0\\
    y_{t,b}\vspace{1mm}\\
    Y_\pm \end{pmatrix}\,,\label{eq:qlYukawas}
\end{align}
with $y_{t,b}$ being the top and bottom Yukawa couplings, respectively, and $Y_\pm = |Y_\pm| e^{i\phi_\pm}$ a complex parameter. 
Stringent constraints from $B_s-\bar{B}_s$ mixing mediated by the neutral $G',Z'$ gauge bosons at tree-level imply that a significant amount of alignment to the down-quark mass basis in the 2-3 sector is phenomenologically required~\cite{Crosas:2022quq}. The dominant breaking of $U(2)_q$ that generates 2-3 mixing in the Cabibbo–Kobayashi–Maskawa (CKM) mixing matrix must therefore be realized in the up-quark sector, corresponding to the limit $Y_- \ll Y_+$. In this case, the contribution of $Y_-$ to the CKM matrix element $V_{cb}$ can be neglected, and we have
\begin{equation}
\label{eq:VcbYpRelation}
|V_{cb}|\,y_t = s_q |Y_+|.
\end{equation}
Working in the limit $Y_- \rightarrow 0$, the leading Yukawa interactions with the Higgs field are those involving $y_t$ and $Y_+$ 
\begin{align}\label{eq:LYukawa}
    \mathcal{L}_Y &\supset s_q Y_+ \bar{q}_L^2 \tilde{H}u_R^3 - y_t\bar{q}_L^3 \tilde{H}u_R^3 - c_q Y_+\bar{Q}_L\tilde{H}u_R^3 
    +s_\ell Y_+^\nu \bar{\ell}_L^2 \tilde{H}\nu_R^3 - y_\nu\bar{\ell}_L^3 \tilde{H}\nu_R^3 - c_\ell Y_+^\nu\bar{L}_L\tilde{H}\nu_R^3 + \mathcal{O}(y_b,y_\tau)\,,
\end{align}
where the couplings $y_\nu$ and $Y_+^\nu$ are defined as\footnote{In the model as written, the $W$-matrix is the only source of quark-lepton splitting and the angle $\chi$ and the phase $\phi_R = -{\rm arg}(y_\tau/y_b)$ are therefore fixed by $y_\tau/y_b$. However, on general grounds, we expect additional sub-leading $SU(4)_h$-breaking corrections such as $\bar\psi_L \Omega_{15} H \psi_R^-$ from integrating-out heavier fields that do not impact the low-energy phenomenology. These corrections can however have a significant impact on the (small) down-type Yukawas, so we treat $\chi$ and $\phi_R$ as free parameters in what follows.}
\begin{equation}
\label{eq:YukawaWMatrix}
\begin{pmatrix}
    y_\nu\vspace{1mm}\\
    Y_+^\nu \end{pmatrix} = W^\dagger 
    \begin{pmatrix}
    y_t \vspace{1mm}\\
    Y_+ \end{pmatrix} \,.
\end{equation}
Finally, the smallness of the tau-neutrino mass is obtained via the addition of a singlet fermion $S_L$ in order to implement the inverse-seesaw mechanism~\cite{Greljo:2018tuh,Fuentes-Martin:2020pww}. Effectively, it interacts with $\nu_R^3$ through the Lagrangian
\begin{equation}
    -\mathcal{L}_{S} = \lambda_R \bar{S}_L\Omega_1 \psi_R^+ + \frac{1}{2}\bar{S}_L \mu S^c_L\,,
    \label{eq:LS}
\end{equation}
resulting in a pseudo-Dirac pair split by the lepton number violating parameter $\mu$. In particular, in the limit $\mu \ll \lambda_R \langle\Omega_1\rangle$, these states have nearly degenerate masses $m_R\approx \lambda_R \langle\Omega_1\rangle \pm \mu/2$ tied to the 4321 breaking scale. After EW symmetry breaking there is also a light, active Majorana state with mass $m_\nu \approx \mu (y_\nu \langle H\rangle/m_R)^2$. In order to obtain small active neutrino masses for $m_R \sim$ TeV and $y_{\nu}\sim \mathcal{O}(1)$, we need to work near the (technically natural) $U(1)_L$ preserving limit $\mu\sim0$. Thus, it is a very good approximation to consider only a single heavy Dirac neutrino of mass $m_R$, and we work in this limit in what follows.~\\

\section{Matching, Running, and Observables in the SMEFT}\label{sec:nlo}
In this section, we discuss the matching of the model introduced in the previous section to the SMEFT. We normalize the Lagrangian as
\begin{equation}
    \cL_{\textrm{SMEFT}} = \sum_{k} \cC_k \cO_k\,,
\end{equation}
where the operators of interest, $\cO_k$, are listed in~\cref{tab:higgsoperators} and~\ref{tab:fourfermionoperators}.
We perform tree-level and one-loop matching, with the following guiding principles:
\begin{itemize}
    \item \textit{Tree-level}. All operators generated by tree-level integration of the heavy fields are included, except those that do not enter EWPO, high-$p_T$ constraints, or $b\to c\tau\nu$ observables, neither directly nor through RGE effects. In particular, integrating out the vector-like quark $Q$ and the right-handed neutrino $\nu_R$ yields some of the Higgs-fermion current operators in~\cref{tab:higgsoperators}, while integrating out the heavy gauge bosons ($U_1$, $G'$, and $Z'$) gives rise to the four-fermion operators in~\cref{tab:fourfermionoperators}.
    \item \textit{One-loop}. One-loop effects are accounted for by taking corrections from $y_t$, $Y_+$, and $g_4$ for the operators entering the electroweak fit directly. In practice, this means that we give the one-loop matching conditions for $\cC_{H\ell}^{(1,3)}$, $\cC_{Hq}^{(1,3)}$, and $\cC_{HD}$, with third-generation flavor indices only. We do not include one-loop matching for $[\mathcal{O}_{Hu}]_{33}$ since it impacts EWPO only via RG mixing into $\cO_{HD}$.  Operators with light fermion fields are always suppressed, e.g. by $s_{q,\ell} \lesssim 0.1$, and have a minimal impact on the electroweak fit. Finally, we include one-loop corrections in $g_4$ for the semi-leptonic operators relevant for $b\to c\tau \nu$ and high-$p_T$ observables.
\end{itemize}

\begin{table}[t]
    \centering
    \renewcommand{\arraystretch}{1.6}
    \setlength{\tabcolsep}{12pt}
    \begin{tabular}{|c|c||c|c|}
        \hline
        $[\cO_{Hu}]_{ij}$ & $(i H^\dagger \overleftrightarrow{D}_\mu H)(\bar{u}_R^i \gamma^\mu u_R^j)$ & $[\cO_{H\ell}^{(1)}]_{\alpha\beta}$ & $(iH^\dagger \overleftrightarrow{D}_\mu H)(\bar{\ell}_L^\alpha \gamma^\mu \ell_L^\beta)$ \\
        $[\cO_{Hq}^{(1)}]_{ij}$ & $(iH^\dagger \overleftrightarrow{D}_\mu H)(\bar{q}_L^i \gamma^\mu q_L^j)$ & $[\cO_{H\ell}^{(3)}]_{\alpha\beta}$ & $(iH^\dagger \overleftrightarrow{D}_\mu^I H)(\bar{\ell}_L^\alpha \gamma^\mu \tau^I \ell_L^\beta)$ \\
        $[\cO_{Hq}^{(3)}]_{ij}$ & $(iH^\dagger \overleftrightarrow{D}_\mu^I H)(\bar{q}_L^i \gamma^\mu \tau^I q_L^j)$ & $\cO_{HD}$ & $|H^\dagger D_\mu H|^2$ \\ \hline
    \end{tabular}
    \caption{Dimension-6 SMEFT operators with the Higgs doublet $H$ generated by the model.}
    \label{tab:higgsoperators}
\end{table}

\begin{table}[t]
    \centering
    \renewcommand{\arraystretch}{1.6}
    \resizebox{\textwidth}{!}{
    \setlength{\tabcolsep}{8pt}
    \begin{tabular}{|c|c||c|c||c|c|}
        \hline
        $[\cO_{\ell q}^{(1)}]_{\alpha\beta ij}$ & $(\bar{\ell}_L^\alpha \gamma^\mu \ell_L^\beta)(\bar{q}_L^i \gamma^\mu q_L^j)$ & $[\cO_{ed}]_{\alpha\beta ij}$ & $(\bar{e}_R^\alpha \gamma^\mu e_R^\beta)(\bar{d}_R^i \gamma^\mu d_R^j)$ & $[\cO_{uu}]_{ijkl}$ & $(u_R^i \gamma^\mu u_R^j)(\bar{u}_R^k \gamma^\mu u_R^l)$ \\
        $[\cO_{\ell q}^{(3)}]_{\alpha\beta ij}$ & $(\bar{\ell}_L^\alpha \gamma^\mu \tau^I \ell_L^\beta)(\bar{q}_L^i \gamma^\mu \tau^I q_L^j)$ & $[\cO_{q q}^{(1)}]_{ijkl}$ & $(\bar{q}_L^i \gamma^\mu q_L^j)(\bar{q}_L^k \gamma^\mu q_L^l)$ & $[\cO_{dd}]_{ijkl}$ & $(d_R^i \gamma^\mu d_R^j)(\bar{d}_R^k \gamma^\mu d_R^l)$ \\
        $[\cO_{\ell edq}]_{\alpha\beta ij}$ & $(\bar\ell_L^{\alpha} e_R^\beta)(\bar d_R^i q_L^j)$ & $[\cO_{q q}^{(3)}]_{ijkl}$ & $(q_L^i \gamma^\mu \tau^I q_L^j)(\bar{q}_L^k \gamma^\mu \tau^I q_L^l)$ & $[\cO_{ud}^{(1)}]_{ijkl}$ & $(u_R^i \gamma^\mu u_R^j)(\bar{d}_R^k \gamma^\mu d_R^l)$ \\
        $[\cO_{\ell u}]_{\alpha\beta ij}$ & $(\bar{\ell}_L^\alpha \gamma^\mu \ell_L^\beta)(\bar{u}_R^i \gamma^\mu u_R^j)$ & $[\cO_{q u}^{(1)}]_{ijkl}$ & $(\bar{q}_L^i \gamma^\mu q_L^j)(\bar{u}_R^k \gamma^\mu u_R^l)$ & $[\cO_{ud}^{(8)}]_{ijkl}$ & $(u_R^i \gamma^\mu T^A u_R^j)(\bar{d}_R^k \gamma^\mu T^A d_R^l)$ \\
        $[\cO_{\ell d}]_{\alpha\beta ij}$ & $(\bar{\ell}_L^\alpha \gamma^\mu \ell_L^\beta)(\bar{d}_R^i \gamma^\mu d_R^j)$ & $[\cO_{q u}^{(8)}]_{ijkl}$ & $(\bar{q}_L^i \gamma^\mu T^A q_L^j)(\bar{u}_R^k \gamma^\mu T^A u_R^l)$ & $[\cO_{\ell\ell}]_{\alpha\beta\gamma\delta}$ & $(\bar{\ell}_L^\alpha \gamma^\mu \ell_L^\beta)(\bar{\ell}_L^\gamma \gamma^\mu \ell_L^\beta)$ \\
        $[\cO_{eq}]_{\alpha\beta ij}$ & $(\bar{e}_R^\alpha \gamma^\mu e_R^\beta)(\bar{q}_L^i \gamma^\mu q_L^j)$ & $[\cO_{qd}^{(1)}]_{ijkl}$ & $(\bar{q}_L^i \gamma^\mu q_L^j)(\bar{d}_R^k \gamma^\mu d_R^l)$ & $[\cO_{\ell e}]_{\alpha\beta\gamma\delta}$ & $(\bar{\ell}_L^\alpha \gamma^\mu \ell_L^\beta)(\bar{e}_R^\gamma \gamma^\mu e_R^\beta)$ \\
        $[\cO_{eu}^{\alpha\beta ij}]_{ijkl}$ & $(\bar{e}_R^\alpha \gamma^\mu e_R^\beta)(\bar{u}_R^i \gamma^\mu u_R^j)$ & $[\cO_{qd}^{(8)}]_{ijkl}$ & $(\bar{q}_L^i \gamma^\mu T^A q_L^j)(\bar{d}_R^k \gamma^\mu T^A d_R^l)$ & $[\cO_{ee}]_{\alpha\beta\gamma\delta}$ & $(\bar{e}_R^\alpha \gamma^\mu e_R^\beta)(\bar{e}_R^\gamma \gamma^\mu e_R^\beta)$ \\
        \hline
    \end{tabular}
    }
    \caption{Dimension-6 SMEFT operators with four fermion fields generated by the model.}
    \label{tab:fourfermionoperators}
\end{table}

\noindent In the following, we give the results for the tree-level matching and the leading running effects. The expressions for the finite parts from the one-loop computation and a more detailed discussion of the matching procedure are given in~\cref{app:AppendixA}.

\subsection{Tree-level matching}
\noindent
All tree-level matching conditions are given in~\cref{tab:TLmatching}.
For example, integrating out the vector-like quark $Q$ gives $[\mathcal{O}_{Hu}]_{33}$, whose RG mixing into the $\cO_{HD}$ operator (proportional to $y_t^2$) provides a shift to the $W$ mass and couplings of the EW gauge bosons to fermions.
Furthermore, integrating out the massive Dirac neutrino gives $[\cC_{H\ell}^{(3)}]_{33}=-[\cC_{H\ell}^{(1)}]_{33}$, resulting in tree-level modifications to $Z$-boson couplings to $\tau$-neutrinos as well as the $W$-boson coupling to $\tau\nu_\tau$.
In the matching of the four-fermion operators, we introduce the effective scale
\begin{align}
    \Lambda_U = \frac{\sqrt{2}\, m_U}{g_4} \,,
    \label{eq:LU}
\end{align}
while the coupling matrix $\beta_L$
(omitting couplings with the vector-like fermions) explicitly reads
\begin{equation}
\label{eq:betaLExpl}
    \beta_L=\begin{pmatrix}
    0& 0& 0 \\
    0 & s_\ell s_q c_\chi & s_q s_\chi \\
    0 & - s_\ell s_\chi & c_\chi
    \end{pmatrix}\,,
\end{equation}
where the fact that the $U_1$ does not interact with the first generation prior to electroweak symmetry breaking is manifest.
Of the four-fermion operators generated by integrating out the $G'$ and $Z'$, we retain only those with purely third-generation flavor indices as they are the only relevant ones for our purposes. In principle, the Wilson coefficients of $[\mathcal{O}_{H\ell}^{(1,3)}]_{22}$ are also generated through the mixing of vector-like leptons with the second-generation leptons
\begin{align}
\label{eq:GFsl}
[\cC_{H\ell}^{(3)}]_{22}^{\rm{tree}} = -[\cC_{H\ell}^{(1)}]_{22}^{\rm{tree}} &= \frac{|Y_+^\nu|^2}{4 m_R^2}\,s_\ell^2
\,,
\end{align}
and the Wilson coefficient $[\cC_{H\ell}^{(3)}]_{22}$ enters the electroweak fit via a modification of the muon decay width that is used to extract the Fermi constant $G_F$. However, including $s_\ell$ as a parameter would open up a new flat direction in the EW fit, requiring an $s_\ell$-sensitive observable to lift. In this work, we set $s_\ell=0$ due to a lack of experimental evidence for large LFU violating effects in the $b\to s\bar \ell \ell$ system.

\begin{table}[]
    \centering
    \resizebox{\textwidth}{!}{
    \renewcommand{\arraystretch}{1.4}
    \setlength{\tabcolsep}{8pt}
    \begin{tabular}{|c|c||c|c||c|c|}
        \hline \rule{0pt}{15pt}
         $\cC_{H\ell}^{(1)}$ & $\displaystyle \frac{|y_\nu|^2}{4 m_R^2}$ &
         $\cC_{qe}$ & $\displaystyle\frac{1}{4 \LU^2}\frac{1}{\xZ}$ & $\cC_{qd}^{(8)}$ & $\displaystyle -\frac{2}{\LU^2} \frac{1}{\xG}$\\[6pt]\hline \rule{0pt}{15pt}
         $\cC_{H\ell}^{(3)}$ & $\displaystyle  -\frac{|y_\nu|^2}{4 m_R^2}$ &
         $\cC_{eu}$ & $\displaystyle \frac{1}{4 \LU^2}\frac{1}{\xZ}$ & $\cC_{uu}$ & $\displaystyle-\frac{1}{3 \LU^2}\left(\frac{1}{\xG}+\frac{1}{8\xZ}\right)$\\[6pt]\hline \rule{0pt}{15pt}
         $\cC_{Hu}$ & $\displaystyle - \frac{|Y_+|^2}{2 m_Q^2}$ &
         $\cC_{ed}$ & $\displaystyle -\frac{1}{\LU^2}\left(1-\frac{1}{4\xZ}\right)$ &$\cC_{dd}$ & $\displaystyle -\frac{1}{3\LU^2}\left(\frac{1}{\xG}+\frac{1}{8\xZ}\right)$\\[6pt]\hline
         \rule{0pt}{15pt}$\cC_{\ell q}^{(1)}$ & $\displaystyle - \frac{1}{2 \Lambda_U^2}\left(c_\chi^2-\frac{1}{2\xZ}\right)$ & 
         $\cC_{qq}^{(1)}$ & $\displaystyle-\frac{1}{12 \LU^2}\left(\frac{1}{\xG}+\frac{1}{2\xZ}\right)$ & $\cC_{ud}^{(1)}$ & $\displaystyle -\frac{1}{12\LU^2} \frac{1}{\xZ}$\\[6pt]\hline \rule{0pt}{16pt}
         $[\cC_{\ell q}^{(3)}]_{\alpha\beta ij}$ & $\displaystyle -\frac{\beta_L^{i\beta}\beta_L^{j\alpha*}}{2 \Lambda_U^2}$ &
         $\cC_{q q}^{(3)}$ & $\displaystyle - \frac{1}{4 \LU^2}\frac{1}{\xG}$ & $\cC_{ud}^{(8)}$ & $\displaystyle -\frac{2}{\LU^2} \frac{1}{\xG}$\\[6pt]\hline \rule{0pt}{15pt}
         $[\cC_{\ell edq}]_{333i}$ & $\displaystyle 2\,\frac{\beta_L^{i3*}e^{i\phi_R}}{\LU^2}$ &
         $\cC_{qu}^{(1)}$ & $\displaystyle - \frac{1}{12 \LU^2}\frac{1}{\xZ}$ & $\cC_{\ell\ell}$ & $\displaystyle - \frac{3}{8\LU^2}\frac{1}{\xZ}$\\[6pt]\hline \rule{0pt}{15pt}
         $\cC_{\ell u}$ & $\displaystyle\frac{1}{4 \LU^2}\frac{1}{\xZ}$ & $\cC_{qu}^{(8)}$ & $\displaystyle - \frac{2}{\LU^2}\frac{1}{\xG}$ & $\cC_{\ell e}$ & $\displaystyle - \frac{3}{4\LU^2}\frac{1}{\xZ}$\\[6pt] \hline \rule{0pt}{15pt}
         $\cC_{\ell d}$ & $\displaystyle \frac{1}{4\Lambda_U^2}\frac{1}{\xZ}$ & $\cC_{qd}^{(1)}$ & $\displaystyle -\frac{1}{12\LU^2} \frac{1}{\xZ}$ &$\cC_{e e}$ & $\displaystyle - \frac{3}{8\LU^2}\frac{1}{\xZ}$\\[6pt] \hline
    \end{tabular}
    }
    \caption{Wilson coefficients generated at tree-level. When flavor indices are omitted, it is assumed that they correspond to purely third-family fermions. We have kept the flavor indices explicit for the Wilson coefficients of semi-leptonic operators that contribute to the charged-current $B$-anomalies. The couplings are defined in Eqs.~\eqref{eq:ULag} and \eqref{eq:LYukawa}, the effective scale $\LU$ in Eq.~\eqref{eq:LU}, while $x_V = m_V^2/m_U^2$ with $V=\Zp,\Gp$ and $m_U$ is the $U_1$ LQ mass.  }
    \label{tab:TLmatching}
\end{table}

\subsection{Leading running effects}
\label{sec:runningEffects}
\noindent
Working with the complete UV model, we are able to compute one-loop corrections arising from the matching conditions at the scale of new physics. The leading effects are encoded in the leading-log (LL) and next-to-leading log (NLL) contributions arising due to the RG evolution of those operators induced at the tree level. In this work, we consider LL running in the three largest SM couplings $y_t$, $g_s$, $g_L$, as well as NLL running in $y_t$ and $g_s$. Here, we summarize the leading running effects, which are due to $y_t$.

\subsubsection{Leading log}
\label{sec:LL}
\noindent
In the following, we report the LL contributions to the Wilson coefficients of the operators affecting EWPO.
Starting with $\cO_{Hq}^{(1)}$, which receives contributions from the running of $\cO_{qq}^{(1)}$, $\cO_{qq}^{(3)}$, $\cO_{qu}^{(1)}$, and $\cO_{Hu}$, the corresponding Wilson coefficient reads
\begin{align} \label{eq:LLHqyt}
    &[\cC_{Hq}^{
    (1)}]_{33}^{\rm{LL}} = \frac{y_t^2}{64\pi^2} \frac{|Y_+|^2}{m_Q^2 }
    \log\left(\frac{\mu^2}{m_Q^2}\right)
    -\frac{y_t^2}{24\pi^2\LU^2}\left[\frac{2}{\xG}\log\left(\frac{\mu^2}{\mG^2}\right)+\frac{1}{16\xZ}\log\left(\frac{\mu^2}{\mZ^2}\right)\right]\,.
\end{align}
By $SU(2)_L$ arguments, the contribution to the Wilson coefficient of the triplet operator can be obtained as
\begin{equation}
    [\cC_{Hq}^{
    (3)}]_{ij}^{\rm{LL}} =\, - \lim_{Y_+\to 0}\,\, [\cC_{Hq}^{
    (1)}]_{ij}^{\rm{LL}}\,.
\end{equation}
Moreover, the running of the semi-leptonic operators $\cO_{\ell q}^{(1,3)}$ results in effects in $\cO_{H\ell}^{(1,3)}$, as already shown in~\cite{Allwicher:2021ndi}. 
In addition, the operators $\cO_{H\ell}^{(1,3)}$ also run into themselves. The complete leading-log result is therefore
\begin{equation}
\label{eq:LLHLyt}
    [\cC_{H\ell}^{(1)}]_{33}^{\rm{LL}} = -  \frac{N_c}{32\pi^2}\frac{y_t^2 c_\chi^2}{\Lambda_U^2} \log\bigg(\frac{\mu^2}{m_U^2}\bigg)
+\frac{N_c}{64 \pi^2} \frac{y_t^2 |y_{\nu}|^2}{m_R^2}
\log\bigg(\frac{\mu^2}{m_R^2}\bigg)
    \,,
\end{equation}
and $[\cC_{H\ell}^{(3)}]_{33}^{\rm{LL}}=-[\cC_{H\ell}^{(1)}]_{33}^{\rm{LL}}$.
Finally, the operator $\cO_{Hu}$ runs into $\cO_{HD}$, resulting in
\begin{equation}
    [\cC_{HD}]^{\rm{LL}} = \frac{N_c }{8\pi^2} \frac{y_t^2 |Y_+|^2}{m_{Q}^2}\log\bigg(\frac{\mu^2}{m_Q^2}\bigg)\,.\label{eq:CHDLL}
\end{equation}

\subsubsection{Next-to-leading log}
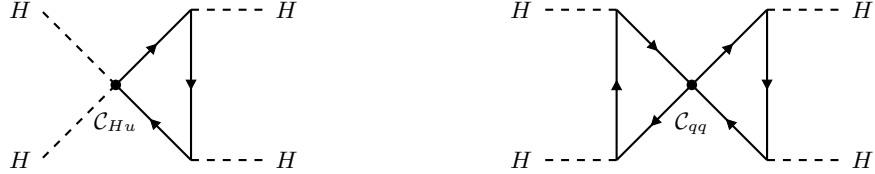
\begin{figure}[t]
    \centering
    \begin{tabular}{c@{\hskip 1cm}c}
    \begin{tikzpicture}[thick,>=stealth]
        \node[wc] at (0,0) {};
        \node[below=7pt] at (0,0) {$\cC_{Hu}$};
        \draw[dashed] (0,0) -- (-1,1) node[left] {$H$};
        \draw[dashed] (0,0) -- (-1,-1) node[left] {$H$};
        \draw[midarrow] (1,-1) -- (0,0);
        \draw[midarrow] (0,0) -- (1,1);
        \draw[midarrow] (1,1) -- (1,-1);
        \draw[dashed] (1,1) -- (2,1) node[right] {$H$};
        \draw[dashed] (1,-1) -- (2,-1) node[right] {$H$};
    \end{tikzpicture}
    & \hspace{15mm}
    \begin{tikzpicture}[thick,>=stealth]
        \node[wc] at (0,0) {};
        \node[below=7pt] at (0,0) {$\cC_{qq}$};
        \draw[midarrow] (-1,1) -- (0,0);
        \draw[midarrow] (0,0) -- (-1,-1);
        \draw[midarrow] (-1,-1) -- (-1,1);
        \draw[midarrow] (0,0) -- (1,1);
        \draw[midarrow] (1,1) -- (1,-1);
        \draw[midarrow] (1,-1) -- (0,0);
        \draw[dashed] (-1,1) -- (-2,1) node[left] {$H$};
        \draw[dashed] (1,1) -- (2,1) node[right] {$H$};
        \draw[dashed] (-1,-1) -- (-2,-1) node[left] {$H$};
        \draw[dashed] (1,-1) -- (2,-1) node[right] {$H$};
    \end{tikzpicture}
    \end{tabular}
    \caption{Diagrams giving the LL (left) and NLL (right) RG running into $\cC_{HD}$. Dotted vertices correspond to insertions of the indicated tree-level WCs (with purely third-family flavor indices), while non-dotted vertices correspond to insertions of $y_t$. In the right diagram, $\cC_{qq}$ should be understood as a sum over $\cC_{qq}^{(1)}$, $\cC_{qq}^{(3)}$, and $\cC_{uu}$, as defined in \cref{tab:TLmatching}.}
    \label{fig:RunningLoops}
\end{figure}
\noindent
We find that the NLL running can give important contributions to the operators relevant for the electroweak fit.
It can be calculated as
\begin{equation}
[\cC]^{\rm NLL} =\frac{1}{2} 
\,\mathcal{A}^2\, [\cC]^{\rm tree}\,\log^2\left(\frac{\mu}{\mu_{0}}\right),
\end{equation}
where $[\cC]$ represents the vector of all dimension-6 Wilson coefficients and $\mathcal{A}$ is the anomalous dimension matrix that can be read from~\cite{Jenkins:2013zja,Jenkins:2013wua,Alonso:2013hga} (see~\cref{app:AppendixB} for more details).
The most important contribution comes from the top Yukawa-induced NLL running of the four-quark operators $\cO_{qq}^{(1)}$, $\cO_{qq}^{(3)}$, and $\cO_{uu}$ generated by tree-level coloron exchange into $\cO_{HD}$, corresponding to the two-loop diagram shown on the right in~\cref{fig:RunningLoops}. This NLL contribution to $\cC_{HD}$ reads
\begin{align}
\label{eq:CHDNLL}
[\cC_{HD}]^{\rm NLL}=
\frac{2 N_c \, y_t^4}{(16\pi^2)^2} \left[(1+2N_c)\cC_{qq}^{(1)} + 3 \cC_{qq}^{(3)} + 2 (1 + N_c) \cC_{uu}\right]
\log^2 \left(\frac{\mu^2}{m_{G^{\prime}}^2}\right) =
-\frac{3y_t^4}{32\pi^4 \Lambda^2_Ux_{G^{\prime}}}\log^2 \left(\frac{\mu^2}{m_{G^{\prime}}^2}\right) \,,
\end{align}
where the WCs are understood to have purely third-family flavor indices. 
To illustrate the potential importance of this effect, we convert the leading and next-to-leading log contributions to $\cC_{HD}$ into shifts in $m_W$~(\cref{eq:WmassExpr}) and plot the result in~\cref{fig:NLLEffect}.
One can see for $\Lambda_U \lesssim 2$ TeV that the NLL running can easily be of the same order or larger than the LL contribution. Additionally, both the LL and NLL contributions to $\cC_{HD}$ have the same sign and thus add coherently.
We include this contribution as well as all NLL running contributions in the couplings $g_s$ and $y_t$ from third-family operators generated at the tree level in the UV in our analysis. The complete expressions for the NLL effects are given in~\cref{app:AppendixBNLL}, and the proper treatment of the top Yukawa $y_t$ is discussed in detail in~\cref{app:AppendixB}.

As a final comment, we note that $\cO_{HD}$ violates custodial symmetry, and therefore its generation should correspond to custodial violations within the model. In the case of the LL contribution, the violation comes from $Y_{-} \ll Y_+$, which is required phenomenologically to pass the bounds on $B_s-\bar{B}_s$ mixing. In the case of the NLL contribution, the custodial symmetry violation is of SM origin, namely $y_b \ll y_t$. It is interesting to note that taking the symmetric limit in both cases is not possible, so these effects cannot be switched off by invoking custodial symmetry.

\begin{figure}[t]
    \centering
    \includegraphics[scale=0.85]{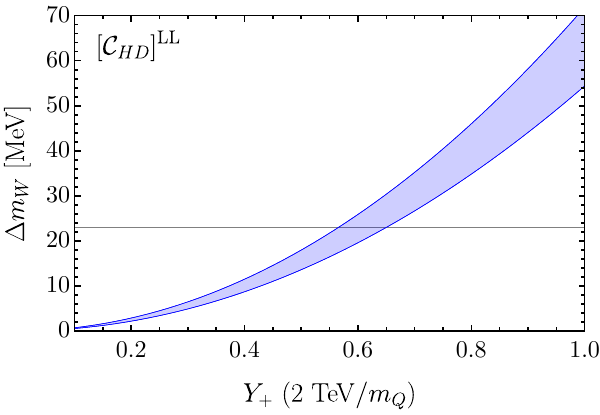} \hspace{5mm}
    \includegraphics[scale=0.835]{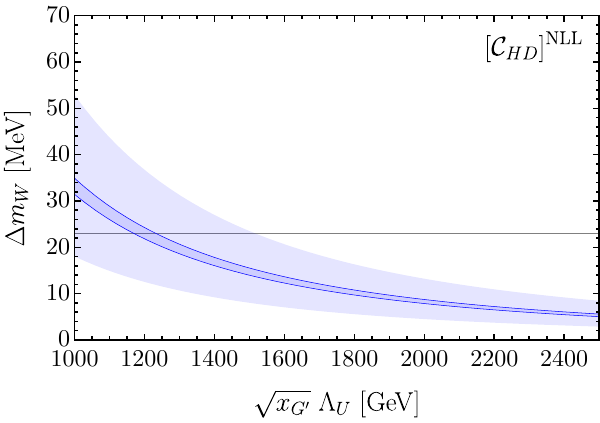}
    \caption{Left: Shift in $m_W$ due to the LL contribution in $\cC_{HD}$. Right: Shift in $m_W$ due to the NLL contribution in $\cC_{HD}$. We are varying the relevant parameters in the ranges $m_Q\in [1.5,3]\,$TeV and  $m_{G^{\prime}}\in [3,3.5]\,$TeV. In the darker blue regions, the IR scale is fixed to $\mu_{\rm EW}=m_t$. In the right plot, the light-blue region shows the impact of varying the IR scale from $m_t/2$ to $2m_t$.
    The gray horizontal line corresponds to the $\Delta m_W$ necessary to accommodate the averaged experimental value $m_W = 80.379$ GeV before CDF II (see \cref{eq:mWold}).}
    \label{fig:NLLEffect}
\end{figure}

\subsection{Observables at one-loop accuracy}
\noindent
Besides the tree-level matching and running effects, we include the one-loop matching in $y_t$, $Y_+$, and $g_4$ of the operators involved in the EW fit, as given in~\cref{app:AppendixA}. Once this is done, for consistency we have to include the one-loop corrections in $y_t$ of the EWPO. These corrections are of the same size as the non-log-enhanced terms of the one-loop matching. If we include these finite pieces from the one-loop matching, the computation of the EWPO at the tree-level in SMEFT would be basis dependent due to the existence of evanescent operators~\cite{Fuentes-Martin:2022vvu}. However, the one-loop calculation of the observables cancels these ambiguities. Moreover, the IR scale dependence of the observables is also canceled at the leading-log order. For all beyond leading-log effects, we choose to fix $\mu_{\rm EW} = m_t$.
Working in the $\{\alpha_{EM},m_Z,G_F\}$ input scheme~\cite{Breso-Pla:2021qoe}, the EWPO can be written in terms of the $W$ and $Z$ bosons vertex corrections and $W$ boson mass shift. In~\cref{app:EWPO} we provide all the expressions for these pseudo-observables, including one-loop corrections in $y_t$.

\section{Phenomenology and Global Fit}\label{sec:pheno}

In the following, we discuss the phenomenology of the model given in~\cref{sec:model} featuring third-family quark-lepton unification. We begin with some general remarks about the expected NP effects in EWPO, breaking them down into flavor non-universal and flavor universal effects:
\begin{itemize}
    \item \emph{Flavor non-universal effects.}
At tree level, the only operators generated by the model that directly affect EWPO are [$\cC_{H\ell}^{(1)}]_{33}$ and $[\cC_{H\ell}^{(3)}]_{33}$, leading to effects in EW gauge boson couplings to third-family leptons. However, because they are generated with the relation $\cC_{H\ell}^{(1)} = - \cC_{H\ell}^{(3)}$, there is no modification to $Z$-boson couplings to tau leptons at tree level. Indeed, only $W\rightarrow \tau\nu_\tau$ and $Z\rightarrow \nu_\tau\nu_\tau$ are affected at tree level, as can be seen in~\cref{app:EWPO}. The sign of the effect is such that existing tensions in $W\rightarrow \tau\nu_\tau$ are worsened, while $Z\rightarrow \nu_\tau\nu_\tau$ is made more compatible with data. At the loop level, this statement remains true when considering only LL running in $y_t$ (see~\cref{eq:LLHLyt}), so the leading breaking to the relation $\cC_{H\ell}^{(1)} = - \cC_{H\ell}^{(3)}$ comes from LL running effects in the $SU(2)_L$ gauge coupling $g_L$. We give the expressions for this effect in~\cref{sec:gLrunning} and take it into account in our analysis. However, the effect in $Z\rightarrow \tau_L\tau_L$ is small and we find no significant constraint from it. Similarly, no modification of $Z$-boson couplings to bottom quarks is generated at tree level. The leading effect in $Z\rightarrow b_L b_L$ comes from the breaking of the relation $\cC_{Hq}^{(1)} = - \cC_{Hq}^{(3)}$ by the $y_t$-induced running of $\cC_{Hu}$ into $\cC_{Hq}^{(1)}$ (\cref{eq:LLHqyt}), LL running in $g_L$, and NLL running in $y_t$ and $g_s$. These contributions are all of a similar size and we take them into account. Again, we find no significant correction to $Z\rightarrow b_L b_L$ vertex. There is also a coloron-induced NLL running effect leading to a sizeable modification of the $Z$-boson coupling to right-handed bottom quarks~\eqref{eq:NLLbR}, but this coupling is poorly constrained and leads to no significant impact on the fit.
\item \emph{Flavor universal effects.}
As discussed in detail in~\cref{sec:runningEffects}, LL and NLL running of tree-level operators induced by the new colored states of the model lead to large flavor universal effects in $\cC_{HD}$ (see~\cref{fig:RunningLoops}). Furthermore, these effects have the same sign and thus add coherently, giving a large universal NP effect that is preferred by the current EW data. In particular, the sign of the effect is such that it increases $m_W$ as well as $Z$-pole observables such as $\Gamma_{Z}$, the effective number of neutrinos $N_{\nu}^{\rm eff}$, and asymmetry observables such as $A_{\ell,q}$ and $A_{b,c}^{\rm FB}$ via flavor universal shifts $\delta^U(T^3, Q, \cC_{HD})$  to the $Z$-boson couplings (shown in~\cref{app:EWPO}). Finally, the flavor non-universal operator $[\cO_{H\ell}^{(3)}]_{33}$ generated at tree level gives flavor-universal contributions to $[\cO_{H\ell}^{(3)}]_{ii}$ and $[\cO_{Hq}^{(3)}]_{ii}$ via LL runnning in $g_L$. While this effect is irrelevant compared to the universal contribution from $\cC_{HD}$, we take it into account via the expressions in~\cref{sec:gLrunning}.
\end{itemize}
As we will see below, this flavor universal effect in $\cC_{HD}$, together with the tree-level non-universal effects in $W\rightarrow \tau\nu_\tau$ and $Z\rightarrow \nu_\tau\nu_\tau$, are dominantly responsible for the behavior of the EW fit. \\

\noindent
We now proceed with a global fit taking into account data from low-energy experiments, EWPO, and neutral current Drell-Yan processes at high-$p_T$. 
In principle, the parameters relevant for the phenomenological discussion are:
\begin{enumerate}
 \item[\bf{I.}]  $SU(4)_h$-{vector mediated}:  
    $\chi$,  $\Lambda_U$, $\phi_R$, $m_U$, $\xG$, $\xZ$,
 \item[\bf{II.}]   {Vector-like fermion mediated}:  
     $Y_+$, $s_q$, $s_\ell$, $m_Q$, $m_R$, $m_L$,
\end{enumerate}
with $\xG = \mG^2/\mU^2$ and $\xZ = \mZ^2/\mU^2$, where $m_U$, $\mG$, and $\mZ$ are the masses of the heavy 4321 gauge bosons $U_1$, $\Gp$, and $\Zp$, respectively. Notice that $s_q$ can be written in terms of $Y_+$ using Eq.~\eqref{eq:VcbYpRelation}, which eliminates it as a free parameter. 
While $Y_+=|Y_+|e^{i\phi_+}$ is in general a complex parameter, a non-vanishing phase would make it necessary to re-phase the SM fields to match the standard CKM phase convention (see Appendix B of~\cite{Crosas:2022quq} for details). This re-phasing would affect the couplings of the $U_1$ LQ to second-family quarks, namely
$(\beta_L)_{2i} \to e^{-i\phi_+}(\beta_L)_{2i}$ in~\cref{eq:betaLExpl}.
To obtain the correct sign and maximize the effect of the LQ on $b\to c\tau\nu$ observables, we fix $\phi_+=0$ and consider $Y_+$ to be real in what follows. Furthermore, we fix $s_{\ell}=0$ according to the discussion below~\cref{eq:GFsl}. As can be seen in~\cref{tab:TLmatching}, all tree-level effects are described by the effective scales
\begin{equation}
\frac{1}{\Lambda_U^2} = \frac{g_4^2}{2m_U^2}\,, \hspace{15mm} \frac{1}{\Lambda_Q^2} = \frac{|Y_+|^2}{2m_Q^2}\,, \hspace{15mm} \frac{1}{\Lambda_R^2} = \frac{|y_{\nu}|^2}{4m_R^2}\,,
\label{eq:effScales}
\end{equation}
up to mass splittings between the 4321 gauge bosons encoded by $\xZ$, $\xG\sim 1$. This is no longer true beyond tree level, where the masses appear alone inside logs and loop functions. Still, the leading behavior is captured by the ratios in~\cref{eq:effScales}, so our strategy will be to fix the masses $m_U$, $\xG$, $\xZ$, $m_L$, and $m_R$ to natural values while allowing $\Lambda_U$, $Y_+$, and $m_Q$ to vary in the fit freely. The reason we do not fix all masses is that, from~\cref{eq:YukawaWMatrix}, the coupling
\begin{align}
    y_\nu = c_\chi y_t - s_\chi Y_+ \,,
    \label{eq:Ynu}
\end{align}
is not an independent parameter. Furthermore, we choose to vary $m_Q$ instead of $m_R$ since the scale $\Lambda_R$ is bounded from below by the process $W\rightarrow \tau\nu$ in the EWPO and $\tau$-decays, so fixing $m_R$ prevents the fit from finding points with very light $m_R$ by fine-tuning the two terms in~\cref{eq:Ynu}.
The benchmark point for the fixed parameters, 
and the variables freely varied in the fit, are
summarized as follows
\begin{enumerate}
 \item[\bf{I.}] Fixed parameters:  
    $m_U=3\,$TeV, $\xZ=1$, $\xG=(3.5/3)^2$, $m_R=1.5\,$TeV, and $m_L=1\,$TeV\,,
 \item[\bf{II.}]  Varied parameters:  
    $\Lambda_U$, $Y_+$, $m_Q$, $\chi$, and $\phi_R$\,.
\end{enumerate}

\subsection{Fit Observables and Methodology}
\label{sec:fitObs}
\noindent
We construct the likelihoods used in the fit by building the $\chi^2$-function, defined as
\begin{equation}
\chi^2 = \sum_{ij}[O_{i,\text{exp}}-O_{i,\text{th}}] (\sigma^{-2})_{ij}[O_{j,\text{exp}}-O_{j,\text{th}}] \,,
\end{equation}
where $\sigma^{-2}$ is the inverse of the covariance matrix. Here, we discuss all observables included in the fit and the corresponding theory predictions. The observables we choose to include are:
\begin{itemize}
    \item \textit{LFU tests in  $b\to c\tau\nu$ transitions.} These include the ratios $R_D$ and $R_{D^*}$, as well as $R_{\Lambda_c}$, defined as
    \begin{align}
        R_{D^{(*)}} = \frac{\cB(B \to D^{(*)} \tau \nu)}{\cB(B \to D^{(*)} \ell \nu)} \,, \qquad
        R_{\Lambda_c} = \frac{\cB(\Lambda_b \to \Lambda_c \tau \nu)}{\cB(\Lambda_b \to \Lambda_c \ell \nu)} \,.
    \end{align}
    The expressions for shifts of these observables from their SM predictions in terms of SMEFT operators are given in~\cref{app:lowenergy}.
    The world averages of $R_{D^{(*)}}$ ratios, including the recent LHCb measurement, read~\cite{HFLAV:2019otj} 
\bea
R_{D^*}^{\rm exp} &=& 0.285 \pm 0.010_{\rm stat} \pm 0.008_{\rm syst}\,, \\
R_{D}^{\rm exp} &=& 0.358 \pm 0.025_{\rm stat} \pm 0.012_{\rm syst}\,, 
\eea
with correlation $\rho = -0.29$. For the SM predictions, we use the HFLAV averages~\cite{HFLAV:2019otj}
\begin{align}
    R_{D}^{\text{SM}} = 0.298(4)\,, \hspace{10mm}
    R_{D^*}^{\text{SM}} = 0.254(5)\,.
\end{align}
Additional discussion about the SM predictions for $R_{D^{(*)}}$ and their uncertainties can be found in~\cite{MILC:2015uhg,Na:2015kha,Bernlochner:2017jka,Gambino:2019sif,Bordone:2019vic,Martinelli:2021onb}.
Regarding $R_{\Lambda_c}$, 
we use the recent LHCb measurement as the input~\cite{LHCb:2022piu} and the result in~\cite{Becirevic:2022bev} as the SM prediction
\begin{equation}
    R^{\rm exp}_{\Lambda_c} = 0.242\pm 0.076\,,\quad R_{\Lambda_c}^{\rm SM} = 0.333(13)\,.
\end{equation}

    \item \textit{LFU tests in $\tau$ decays}. These are expressed through the ratios
    \begin{align}
        \left(\frac{g_\tau}{g_{\mu(e)}}\right)_\ell &= \left[\frac{\cB(\tau \to e(\mu) \nu\bar\nu)/\cB(\tau \to e(\mu) \nu\bar\nu)_{\rm SM}}{\cB(\mu\to e\nu\bar\nu)/\cB(\mu\to e\nu\bar\nu)_{\rm SM}}\right]^\frac{1}{2} \\
        \left(\frac{g_\tau}{g_{\mu}}\right)_\pi &= \left[\frac{\cB(\tau \to \pi \nu)/\cB(\tau \to \pi \nu)_{\rm SM}}{\cB(\pi\to \mu\bar\nu)/\cB(\pi \to \mu\bar\nu)_{\rm SM}}\right]^\frac{1}{2} \\
        \left(\frac{g_\tau}{g_{\mu}}\right)_K &= \left[\frac{\cB(\tau \to K \nu)/\cB(\tau \to K \nu)_{\rm SM}}{\cB(K \to \mu\bar\nu)/\cB(K \to \mu\bar\nu)_{\rm SM}}\right]^\frac{1}{2} \,,
    \end{align}
    expected to be equal to one in the SM by definition. We use the experimental average from \cite{Cornella:2021sby}
    \begin{align}
        \left(\frac{g_\tau}{g_{e,\mu}}\right)_{\ell,\pi,K} = 1.0012 \pm 0.0012 \,.
    \end{align}
    The theory predictions for the $\tau$ LFU tests can be found in~\cref{app:lowenergy}.
    \item \textit{Electroweak precision observables}. At the $Z$- and $W$-pole, we include all the observables listed in Tables 1 and 2 of Ref.~\cite{Breso-Pla:2021qoe} using the same SM theory predictions, and we follow the same input scheme. For the NP theory predictions, we parameterize deviations from the SM due to NP effects via shifts in the $W$ and $Z$ boson couplings, $\delta g^{W,Z}_k$, and the shift in the $W$-boson mass $\delta m_W$. To simplify notation, we include $\delta m_W$ in the vector $\delta g^{W,Z}_k$ and define the theory prediction for a given EWPO as
    \begin{equation}
    O_{i,\text{th}} = O_{i,\text{SM}} + \sum_{k} \alpha_{ik} \delta g^{W,Z}_k \,.
    \end{equation}
    The theory predictions for the $\delta g^{Z,W}_k$ in terms of the SMEFT Wilson coefficients at one-loop in $y_t$, as well as the map between the EWPO and the $\delta g^{Z,W}_k$ that determines the $\alpha_{ik}$, are given in~\cref{app:EWPO}.

    We will discuss and compare the fit in the two scenarios, one with and one without the latest CDF II result on $m_W$ \cite{CDF:2022hxs}. In the former case, given the statistical incompatibility of the CDF II measurement with all the previous determinations, a strategy needs to be defined in order to be able to perform a combination. 
Our choice is to assume that all experimental errors have been underestimated.
Penalizing all measurements ``democratically", and inflating the errors until the $\chi^2$ per degree of freedom is equal to one, one finds $m_W = 80.410 \pm 0.015 \, \text{GeV}$ for the averaged $W$ mass. Therefore, we consider two scenarios with different input values for $m_W$
    \begin{align}
    m_W^{\rm old} &= 80.379 \pm 0.012 \, \text{GeV} \hspace{20mm} \text{without CDF II}\,, \label{eq:mWold} \\
    m_W^{\rm new} &= 80.410 \pm 0.015 \, \text{GeV} \hspace{20mm} \text{with CDF II}\,.
    \end{align}
    \item \textit{High-$p_T$ constraints}. We consider the tail of $pp \to \tau\tau$ distributions obtained at the LHC experiments. The likelihood is obtained using {\tt HighPT} \cite{Allwicher:2022mcg}, and for the theory input we use the semi-leptonic SMEFT coefficients computed at NLO in $g_4$, as discussed in~\cref{app:RWC@1loop}. The likelihood is constructed using data from ATLAS~\cite{ATLAS:2020zms}. However, it is interesting to consider also the recent search by CMS~\cite{CMS:2022goy} which shows a small excess, leading to a weaker bound on the NP scale. As done in Ref.~\cite{Aebischer:2022oqe}, we rescale the likelihood obtained from {\tt HighPT} to match the NLO predictions derived in Ref.~\cite{Haisch:2022afh} for the ATLAS and CMS searches in the b-tag channel.
    In addition, a lower bound on the vector-like quark mass $m_Q$ is implemented by introducing a term in the likelihood of the form
    \begin{align}
        \Delta \chi^2 = 4  \left(\frac{m_Q^{\rm bound}}{m_Q}\right)^2 \,,
    \end{align}
    where $m_{Q}^{\rm bound} = 1.5 \, \text{TeV}$ is the lower bound on the mass at 95\% CL~\cite{CMS:2022fck} from direct searches for $SU(2)_L$-doublet vector-like quarks pair-produced via QCD and decaying dominantly to $Ht$ and $Zt$, as we expect in this model.
\end{itemize}

\subsection{Fit Results}
\noindent
As discussed in~\cref{sec:fitObs}, we have the likelihoods $\chi^2_{b\rightarrow c\tau\nu}$, $\chi^2_{\rm EWPO}$, $\chi^2_{\tau\text{-LFU}}$, and $\chi^2_{\text{high-}p_T}$. We define the global likelihood simply as the sum of the individual likelihoods
\begin{equation}
\chi^2 = \chi^2_{b\rightarrow c\tau\nu} + \chi^2_{\rm EWPO} + \chi^2_{\tau\text{-LFU}}+ \chi^2_{\text{high-}p_T} \,.
\end{equation}
Fixing the parameters as discussed in the beginning of~\cref{sec:pheno}, the global likelihood is a function of 5-parameters, namely $\chi^2(\Lambda_U, Y_+, m_Q, \chi, \phi_R)$. However, as can be seen in~\cref{fig:chiPlot}, when choosing $m_W^{\rm new}$ the global $\chi^2$-function is very flat for $\chi \in [35^{\circ},80^{\circ}]$, where it changes by less than 2 units along the red line giving the optimal value of $\Lambda_U$. However, the optimal value of $Y_+$ (dashed black contours) is not flat in this range. Instead, it decreases from around 1 for $\chi = 35\degree$ to around 0.4 for $\chi = 80\degree$. Generally speaking, larger values of $Y_+$ favor the EWPO which calls for non-zero $\cC_{HD}\propto |Y_+|^{2}$, while smaller values of $Y_+$ favor the $b\rightarrow c\tau\nu$ observables, which call for larger $s_q \propto |Y_+|^{-1}$. Due to the flatness of the $\chi^2$-function between $35^{\circ}$ and $80^{\circ}$, we choose to fix $\chi = 60^{\circ}$ for everything that follows, to allow for an intermediate value of $Y_+ \sim 0.6$ that strikes a compromise between EWPO and $b\rightarrow c\tau\nu$ observables while making essentially no difference in the goodness of the fit. We have explicitly checked that this choice for $\chi$ is also good for the fit using $m_W^{\rm old}$.

\begin{figure}
    \centering
    \includegraphics[scale=1.1]{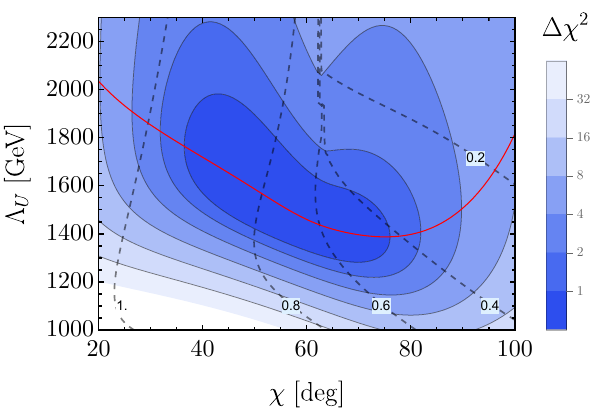}
    \caption{The global likelihood $\Delta\chi^2$ (blue contours) relative to the minimal value as a function of the angle $\chi$ and NP scale $\Lambda_U$, finding the value of $Y_+$ that minimizes the $\chi^2$ function for each point. The optimal values of $Y_+$ are shown by the black dashed contour lines. The red line shows the value of $\Lambda_U$ that minimizes the $\chi^2$ for each value of the angle $\chi$. All other parameters are fixed or set to their best-fit values for $m_W^{\rm new}$ given in~\cref{tab:fitresults}. }
    \label{fig:chiPlot}
\end{figure}

\noindent
Having fixed the angle $\chi=60^{\circ}$, we minimize the global $\chi^2$-function with respect to the remaining 4 parameters and report their best-fit values and $1\sigma$-confidence intervals in~\cref{tab:fitresults}. The best-fit point (BFP) corresponds to $\Delta \chi^2 = \chi^2_{\rm SM} - \chi^2_{\rm BFP} = 12.3~(15.4)$ in the case of $m_W^{\rm old}~(m_W^{\rm new})$, indicating a significant improvement over the SM hypothesis.\footnote{Assuming 4 degrees of freedom
as in  Table~\ref{tab:fitresults}, this change in $\chi^2$ corresponds to a $2.4\sigma~(2.9\sigma)$ improvement over the SM hypothesis. However, the effective number of degrees of freedom is likely to be smaller than 4 (and the significance therefore higher) given the poor sensitivity of the fit to some of the free parameters. On the other hand, we note that the selection of observables we have considered is somewhat biased by the choice of the model. 
}
Next, we show the $1\sigma$ and $2\sigma$ preferred regions in the $\Lambda_U$ vs. $Y_{+}$ plane in~\cref{fig:fit1} for the $b\rightarrow c\tau\nu$ observables (blue), EWPO (red), as well as the global preferred regions including all observables (blue lines without filling). The region below the dashed black line is excluded by CMS $pp\to\tau\tau$ data, while the gray shaded region is excluded by $\tau$-LFU tests, both at 95\% CL. In addition, the explicit relation between the couplings $Y_+$ and $y_\nu$, fixed by Eq.~\eqref{eq:Ynu}, allows us to show the preferred regions for $y_\nu$ (upper horizontal axis in~\cref{fig:fit1}) simultaneously. Importantly, the parameters that are not varied in the plots are fixed to their global best-fit points, except in the case of $m_W^{\rm old}$ where we fix $m_Q =3.2$ TeV. This is due to the insensitivity of the fit to $m_Q$ when $m_Q \gtrsim 2.3$ TeV, so we simply choose a theoretically reasonable value in the $m_W^{\rm old}$ case.
We now offer some general conclusions from the results of the fit.

\begin{table}[t]
    \centering
    \begin{tabular}{c c}
        \begin{tabular}{c|c|c}
         Parameter & Best-fit point & 1$\sigma$ interval \\\hline\hline
         $\Lambda_U$ & 1.61~{\rm TeV} & [1.46,\,1.86]~{\rm TeV} \\
         $m_Q$ & $m_Q \to \infty$ & [2.31,\,$\infty$)~{\rm TeV} \\
         $Y_+$ & 0.36 & [0.26,\,0.56] \\
         $\phi_R$ & 180~{\rm deg} & [127,\,233]~{\rm deg} \\
         $\chi$ & 60~{\rm deg} & {\rm fixed} \\
    \end{tabular}
    \hspace{15mm} &
    \begin{tabular}{c|c|c}
         Parameter & Best-fit point & 1$\sigma$ interval \\\hline\hline
         $\Lambda_U$ & 1.46~{\rm TeV} & [1.32,\,1.68]~{\rm TeV} \\
         $m_Q$ & 2.08~{\rm TeV} & [1.43,\,4.72]~{\rm TeV} \\
         $Y_+$ & 0.65 & [0.43,\,0.83] \\
         $\phi_R$ & 180~{\rm deg} & [135,\,225]~{\rm deg} \\
         $\chi$ & 60~{\rm deg} & {\rm fixed} \\
    \end{tabular}
    \end{tabular}
    \caption{Best-fit point and 1$\sigma$ ranges for the parameters varied in the fit. Left: Using $m_W^{\rm old}$ as an input in the EWPO. Right: Using $m_W^{\rm new}$ as an input in the EWPO. The best-fit point of our NP model corresponds to $\Delta \chi^2 = \chi^2_{\rm SM} - \chi^2_{\rm BFP} = 12.3~(15.4)$ in the case of $m_W^{\rm old}~(m_W^{\rm new})$  over the SM hypothesis.}
    \label{tab:fitresults}
\end{table}
\begin{figure}[]
    \centering
    \begin{tabular}{c c}
    \includegraphics[width=0.5\textwidth]{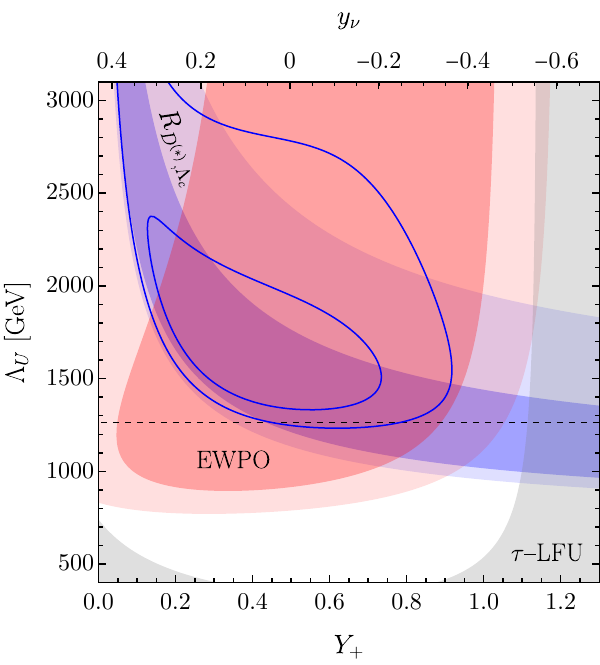} & 
    \includegraphics[width=0.5\textwidth]{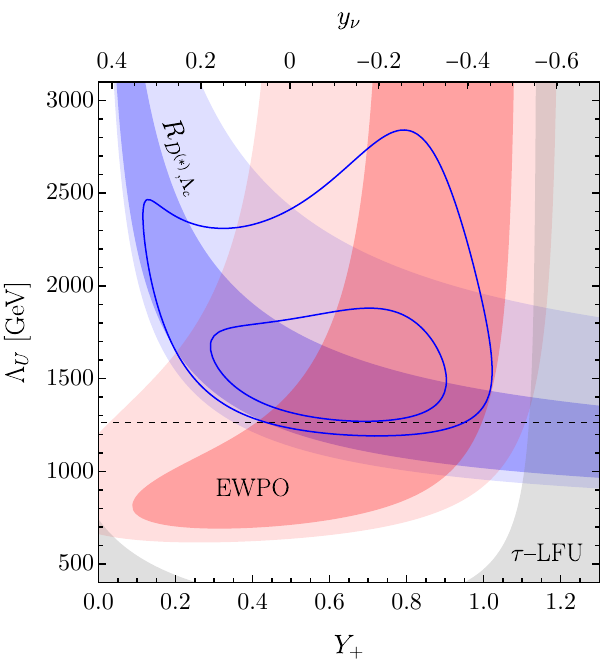}
    \end{tabular}
    \caption{Left: Preferred regions using $m_W^{\rm old}$ in the EWPO. Right: Preferred regions using $m_W^{\rm new}$ in the EWPO. Blue region: $1$ and $2\sigma$ regions preferred by a fit to the low-energy $R_{D}$, $R_{D^{*}}$, and $R_{\Lambda_c}$ observables. Red region: $1$ and $2\sigma$ regions preferred by a fit to EWPO. Blue lines: $1$ and $2\sigma$ regions preferred by a global fit including all observables. Dashed black line: The region below is excluded by the CMS $pp\rightarrow \tau\tau$ search with $b$-tagging at 95\% CL. Gray region: Excluded by $\tau$-LFU tests at 95\% CL. Parameters not varied in the plots are fixed to their global best-fit values, except in the case of $m_W^{\rm old}$ where we fix $m_Q =3.2$ TeV.}
    \label{fig:fit1}
\end{figure}

\noindent
In the case of $b\to c \tau\nu$ transition observables, we remark that the preferred region allows for larger $\Lambda_U$ values as $|Y_+|$ gets smaller. This behavior is mainly dictated by the fact that LFU ratios $R_{D^{(*)}}$ and $R_{\Lambda_c}$ receive contributions proportional to $\beta^{23}_L = s_q s_\chi$, where $s_q = V_{cb} \, y_t/|Y_+|$. Smaller $Y_+$, therefore, increases this contribution, allowing for a higher effective scale $\LU$ to achieve the same size NP effect. Moreover, as shown in~\cite{Crosas:2022quq}, $K$--$\bar{K}$ mixing and measurements of $\mathcal{B}(K^+ \rightarrow \pi^+ \nu\bar\nu)$ limit the magnitude of $s_q \lesssim 0.1$, thus requiring larger values of $|Y_+| \gsim 0.35$. On the other hand, the $1\sigma$ confidence interval for $Y_+$ translates to $s_q \in [0.04,0.08]$. It is very interesting that the results of this fit, which does not include kaon observables, are well compatible with the bound on the leading $U(2)_q$ breaking parameter $s_q$ found in Ref.~\cite{Crosas:2022quq}.
Finally, moving in the direction of larger $|Y_+|$ (smaller $s_q$), we observe the simultaneous decrease in the value of $\LU$ necessary to accommodate the $b\to c \tau\nu$ data as expected.
\\

\noindent
Regarding $\tau$-LFU tests, they exclude $|y_\nu| \gtrsim 0.5$ at 95\% CL for almost all values of $\Lambda_U$, with the bound becoming somewhat stronger for very low values (around 500 GeV) of $\Lambda_U$.
This is due to constructive addition between the tree-level ($\propto -|y_\nu|^2/m_R^2$) and one-loop ($\propto c_\chi^2 \Lambda_U^{-2} \log m_t^2/\Lambda_U^2$) contributions to the operator $\cC_{H\ell}^{(3)}$ (see also the discussion below).
These statements hold where the mass of the right-handed neutrino has been fixed to $m_R = 1.5$ TeV. At the best-fit point, we have $y_\nu = -0.13$, and the $1\sigma$ confidence interval is  $y_\nu \in [-0.29,0.06]$. This corresponds to mixing angles $\theta_\tau^2 \equiv |y_\nu|^2 v^2 / (2m_R^2)$ in the range $\theta_\tau^2 \in [5\times 10^{-5},10^{-3}]$, where $v = 246$ GeV. As a consistency check for the fit, these values are well below the bound obtained from EWPO in the literature of $\theta_\tau^2 < 5.3\times 10^{-3}$~\cite{Antusch:2014woa,Antusch:2015mia}. Switching on only $\cC_{H\ell}^{(1)}= -\cC_{H\ell}^{(3)} =  \theta_{\tau}^2/(2v^2)$, we find a similar but somewhat stronger bound of $\theta_\tau^2 < 2.3\times 10^{-3}$ at 95\% CL when combining the likelihoods from $\tau$-LFU tests and EWPO that neatly explains the preferred range on $|y_{\nu}|$ from the fit.  We note that smaller values of $m_R$ will lead to a tighter upper bound on $|y_\nu|$, while the bound gets relaxed in the large $m_R$ limit. However, since $m_R \propto \langle \Omega_1 \rangle$, it cannot be fully decoupled without simultaneously decoupling the $U_1$ LQ. Therefore, when the scale of the model is low enough to address the charged-current $B$-anomalies, it always predicts sizeable unitarity violations, $\theta_\tau$, in the active neutrino mixing matrix unless $y_{\nu}$ is tuned small, as first pointed out in~\cite{Greljo:2018tuh}.
\\\\

\noindent
Coming next to the EWPO, one can observe that:
\begin{itemize}
    \item[i)] When $m_W^{\rm old}$ is used as an EWPO (left panel in~\cref{fig:fit1}), the fit generally prefers lower values of $|Y_+|$ and $\Lambda_U \gtrsim 1$ TeV, i.e. those regions in which $m_W$ does not receive a large shift. This is a consequence of the fact that $\delta m_W \sim |Y_+|^2$, from the LL contribution to $\cC_{HD}$, as can be seen in~\cref{eq:CHDLL}. Additionally, there is the NLL contribution given in Eq.~\eqref{eq:CHDNLL} that is proportional to $\Lambda_U^{-2}$ and adds constructively with the LL contribution. It is interesting that the lower bound of $\Lambda_U \gtrsim 1$ TeV, coming dominantly from the NLL contribution to $\cC_{HD}$, is stronger than that of $\tau$-LFU tests when $\chi$ is fixed to $60^{\circ}$.
    \item[ii)] When $m_W^{\rm new}$ is used as an EWPO (right panel in~\cref{fig:fit1}), the EW fit prefers to have larger $|\cC_{HD}|$, reflected in the fact that larger values of $|Y_+|$ are now preferred by EW precision data for $\Lambda_U \gtrsim 1.5$ TeV, where the NLL effect is less relevant (as shown in~\cref{fig:NLLEffect}).  Furthermore, we note the preferred region also stretches to smaller values of $|Y_+|$ and $\Lambda_U$, $|Y_+|\in[0.2,0.6]$ and $\Lambda_U < 1.5$ TeV. This is due to the NLL contributions to $\cC_{HD}$ from the four-quark operators induced by the coloron exchange at tree-level,~\cref{eq:CHDNLL}. This effect is inversely proportional to $\Lambda_U^2$, and adds constructively with the NLL contribution, allowing a larger $W$ mass to be accommodated even for small $|Y_+|$, provided $\Lambda_U$ is small enough. 
\end{itemize}

\noindent
Finally, we note that for fixed values of the angle $\chi$, the high-$p_T$ constraints are sensitive only to the value of $\Lambda_U$. The black dashed line gives the 95\% CL exclusion limit from CMS, which is providing by far the most stringent lower bound on the NP scale $\Lambda_U \gtrsim 1.25$ TeV. Similarly, the 95\% CL exclusion limit from the ATLAS data is $\Lambda_U \gtrsim 1.6$ TeV. However, due to the fact that ATLAS currently has an under-fluctuation in the data and is therefore setting more stringent limits than expected (and also does not yet have a dedicated non-resonant search in the $pp\rightarrow \tau\tau$ channel), we consider the CMS bound of $\Lambda_U \gtrsim 1.25$ TeV to be the current hard lower bound on the scale. Therefore, in the global fit, we have used the CMS likelihood for $\chi^2_{\text{high-}p_T}$. Finally, it is worth mentioning that the right-handed couplings of the $U_1$ LQ to the SM fermions can be reduced in less minimal versions of the model, e.g. via mixing with extra vector-like fermions. In such a case, the bound from high-$p_T$ constraints would be reduced, decreasing the tension with the current ATLAS data~\cite{Aebischer:2022oqe}.}

\noindent
\subsection{Selected Observables}
\begin{table}[t]
    \centering
    \renewcommand{\arraystretch}{1.4}
    \begin{tabular}{c c}
    \begin{tabular}{c|c|c|c}
         Observable & $P_{\rm SM}$ & $P_{\rm BFP}$ & ${|P_{\rm BFP}|-|P_{\rm SM}|}$ \\\hline\hline
         $R_{D^*}$ &-2.42 & -1.46& -0.96 \\
         $R_{D}$ &-2.16 & 0.27 & -1.89 \\
         $R_{\Lambda_c}$ & 1.20 & 1.63 & 0.43 \\
         $\left(\frac{g_\tau}{g_{e,\mu}}\right)_{\ell,\pi,K}$ &-1.00 & -1.16& 0.16 \\
        \hline
        $m_{W}^{\rm old}$ &-1.92 & -0.84& -1.08 \\
        $A_{e}$ &-2.19 & -1.91& -0.28 \\
        $\Gamma_{Z}$ &-0.61 & -0.37& -0.24 \\
        $A_{b}^{\rm FB}$ & 2.25 & 2.52 & 0.27 \\
        $A_{\tau}$ & 0.72 & 0.88 & 0.16 \\
    \end{tabular} \hspace{15mm} & 
    \begin{tabular}{c|c|c|c}
         Observable & $P_{\rm SM}$ & $P_{\rm BFP}$ & ${|P_{\rm BFP}|-|P_{\rm SM}|}$ \\\hline\hline
         $R_{D^*}$ &-2.42 & -1.65& -0.77 \\
         $R_{D}$ &-2.16 & -0.18& -1.98 \\
         $R_{\Lambda_c}$ & 1.20 & 1.55 & 0.35 \\
         $\left(\frac{g_\tau}{g_{e,\mu}}\right)_{\ell,\pi,K}$ &-1.00 & -1.15& 0.15 \\
        \hline
        $m_{W}^{\rm new}$ &-3.60 & -1.77& -1.83 \\
        $A_{e}$ &-2.19 & -1.59& -0.60 \\
        $\Gamma_{Z}$ &-0.61 & -0.02& -0.59 \\
        $A_{b}^{\rm FB}$ & 2.25 & 2.82 & 0.57 \\
        $A_{\tau}$ & 0.72 & 1.04 & 0.32 \\
    \end{tabular}
    \end{tabular}
    \caption{Pull of the LFU observables together with the 5 EWPO with the most influence on the global fit. Left: Observable pulls using $m_W^{\rm old}$ in the EWPO. Right: Observable pulls using $m_W^{\rm new}$ in the EWPO.  The final column is negative (positive) if the BFP of the NP model reduces (increases) the tension with the experiment compared to the SM for a given observable. The NP pull $P_{\rm BFP}$ is defined using the values in~\cref{tab:fitresults}, except in the $m_W^{\rm old}$ case, where we fix $m_Q =3.2$ TeV.}
    \label{tab:obsPulls}
\end{table}

\begin{figure}[t]
    \centering
    \resizebox{0.9\textwidth}{!}{
    \begin{tabular}{cccc}
        (a) & \raisebox{-\height}{\includegraphics[height=0.2\textheight]{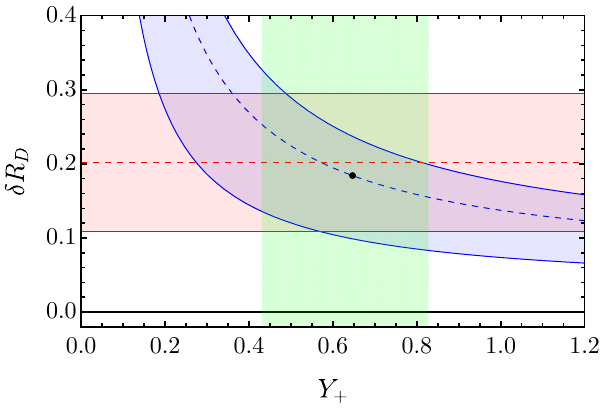}} & (b) & \raisebox{-\height}{\includegraphics[height=0.2\textheight]{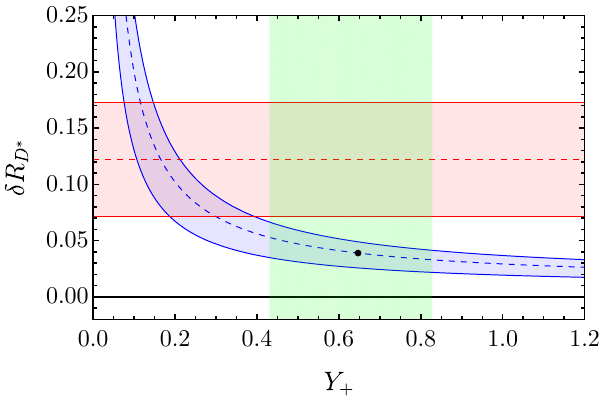}} \\[2.2cm]
        (c) & \raisebox{-\height}{\includegraphics[height=0.2\textheight]{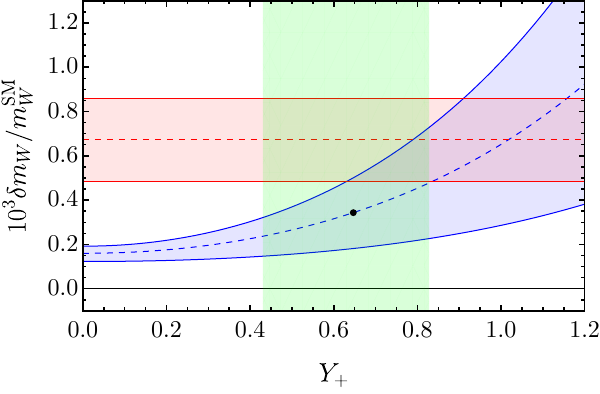}} & (d) & \raisebox{-\height}{\includegraphics[height=0.2\textheight]{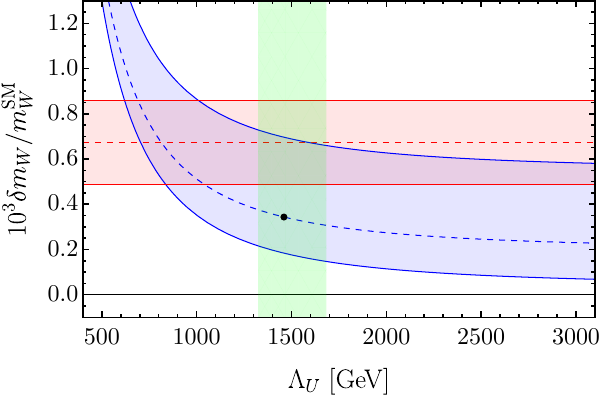}} \\
        (e) & \raisebox{-\height}{\includegraphics[height=0.2\textheight]{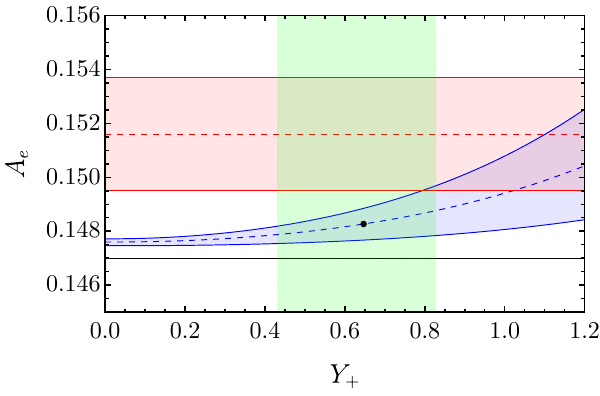}} & (f) & \raisebox{-\height}{\includegraphics[height=0.2\textheight]{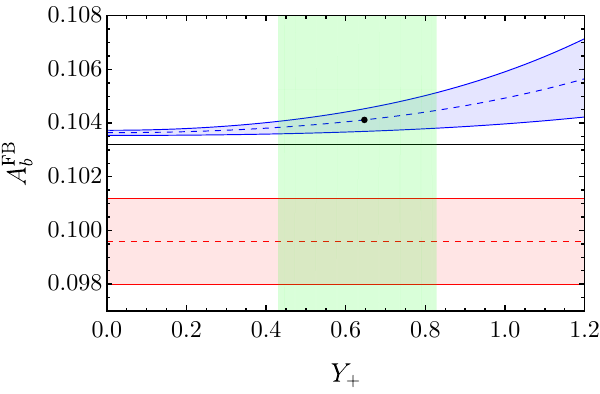}} \\
        (g) & \raisebox{-\height}{\includegraphics[height=0.2\textheight]{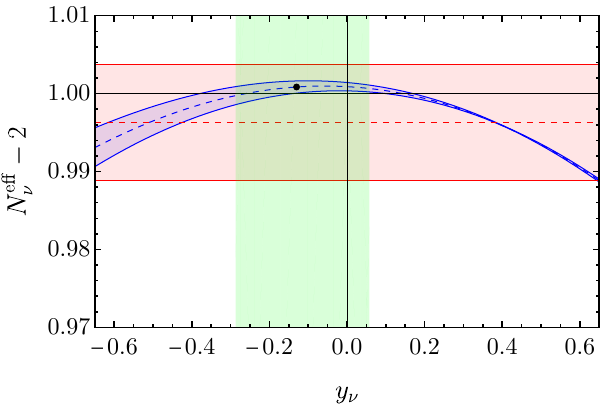}} & (h) & \raisebox{-\height}{\includegraphics[height=0.2\textheight]{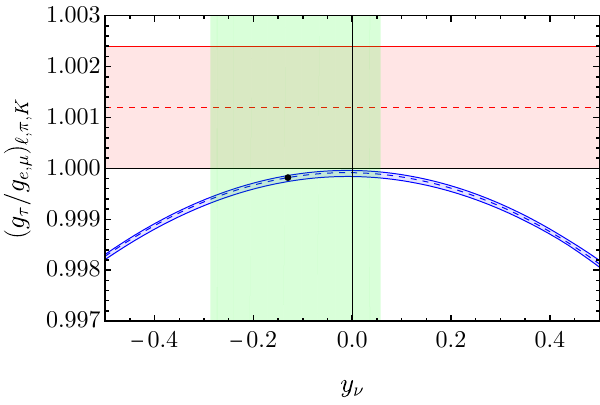}} \\
    \end{tabular}
    }
    \caption{
        The blue (red) bands give the theory ($\pm 1\sigma$ experimental) determinations for the different observables as a function of the respective model parameter.
        The boundaries of the theory band are determined by varying all other parameters within their $1\sigma$ confidence intervals.
        The black dots indicate the global best-fit points, and the green bands correspond to the $1\sigma$ confidence interval of the NP parameter being plotted. In all cases, we use $m_W^{\rm new}$ as an input in the EWPO. The overlap of the blue and green regions should therefore be considered as reasonably achievable within the NP model. Finally, the SM theory predictions are indicated by the solid black line.
    }
    \label{fig:obsPlots}
\end{figure}

\noindent
Having discussed the general behavior of the fit, it is interesting to have a look at the observables that dominantly drive the fit. 
We define the pull of a particular observable $O$ computed in a theoretical model M to be
\begin{equation}
P_{\text{M}}  = \frac{O_{\text{M}}-O_{\text{exp}}}{\sigma_{\rm exp}} \,,
\end{equation}
where $\sigma_{\rm exp}$ is the experimental uncertainty for a given measurement $O_{\text{exp}}$. In~\cref{tab:obsPulls}, we show the highest pulling observables for the best-fit point (BFP) of the NP model as well as the same observables within the SM. Furthermore, in~\cref{fig:obsPlots} we compare the NP model predictions using $m_W^{\rm new}$ (blue regions) with the current experimental determinations for selected individual observables (red regions). We also show the $1\sigma$ interval for the model parameter being plotted in green, so the overlap of the blue and green regions should be considered as reasonably obtainable within the model. We can make the following observations:
\begin{itemize}
    \item $R_D$/$R_{D^*}$ [Figs.~\ref{fig:obsPlots}(a) and~\ref{fig:obsPlots}(b)]. At the best-fit point, we have $\delta R_{D} = 0.18$ and $\delta R_{D^*} = 0.04$, so the model gives a very good fit for $R_D$, while lying slightly below the central experimental value ($\approx 1.7\sigma$) for $R_{D^*}$. The reason why a large contribution to $R_D$ can be obtained even for relatively high $\Lambda_U$ is due to the right-handed $b$-$\tau$ coupling of the $U_1$ LQ in this model. As the fit prefers $\phi_R = \pi$, a large constructive scalar contribution to $R_D$ is generated via $[\cC_{\ell e d q}]_{3332}$ (see \ref{eq:RD}).  In general, as argued before, smaller values of $|Y_+|$ or $\Lambda_U$ are required to increase the contribution to both ratios, which then increases the tension with EWPO or high-$p_T$, respectively.
    \item $m_W$ [Figs.~\ref{fig:obsPlots}(c) and~\ref{fig:obsPlots}(d)]. The opposite is true for $\delta m_W$, where one can see in \cref{fig:obsPlots}(c) that larger values of $|Y_+|$ give a larger NP contribution via LL running into $\cC_{HD}$. In Fig.~\ref{fig:obsPlots}(d), which shows $\delta m_W$ as a function of $\Lambda_U$, one can see the impact of NLL running into $\cC_{HD}$ from coloron exchange, giving a large effect in $m_W$ for small values of $\Lambda_U$. The small $\Lambda_U <1.25$ TeV region is, however, strongly in tension with $pp\rightarrow \tau\tau$ data. We note that the central value of $m_W^{\rm new}$ lies within the overlap of the green and blue regions, and is therefore achievable within the model.
    \item $A_e$ and $A_b^{\rm FB}$ [Figs.~\ref{fig:obsPlots}(e) and~\ref{fig:obsPlots}(f)]. The model predicts an increase with respect to the SM in both cases, again coming from the universal contribution via $\cC_{HD}$. The measurements, however, go in opposite directions, forcing the fit to compromise in order to not significantly increase the pre-existing tension in $A_b^{\rm FB}$.
    \item $N_{\rm eff}$ and $\tau$-LFU [Figs.~\ref{fig:obsPlots}(g) and~\ref{fig:obsPlots}(h)]. The same flavor non-universal NP effects appear in both observables, predicting in general a decrease from the SM expectation. Also here, the experimental central values lie in opposite directions, with NP effects causing more tension with $\tau$-LFU measurements. As large $Y_+$ corresponds to negative $y_\nu$, the asymmetric shape of the $N_{\rm eff}$ prediction for negative values of $y_\nu$ stems from universal contributions to the $Z$-boson couplings to neutrinos via $\cC_{HD}$. Such universal shifts cancel out in the LFU ratios and are therefore not seen in~\cref{fig:obsPlots}(h), where the model always increases the tension with data.
\end{itemize}

\section{Conclusions}
\noindent

We have studied for the first time the phenomenological implications of
a UV complete model featuring third-family quark-lepton unification for 
electroweak precision tests. This was done by matching the
UV model (defined in~\cref{sec:model})
to the SMEFT at one loop, paired with a one-loop computation of the EWPO within the SMEFT (which is necessary to consistently capture all finite parts). We find good agreement between the model predictions for the EWPO and the experimental results, with an overall improvement of the EW fit with respect to the SM. 

At tree level, the most sizeable corrections come from integrating out a heavy pseudo-Dirac singlet necessary to account for acceptable neutrino masses within the model. The primary impact of this state on EWPO and $\tau$-LFU tests is to increase the pre-existing tension in the $W$-boson coupling to third-family leptons.  However, one of the most important results of our analysis is the 
demonstration that higher-order effects  play a key role. They
give rise to a qualitatively different behavior of the fit compared to the inclusion of tree-level effects only,  and break flat directions in the parameter space. In this respect, many of the results we have derived, from the matching to the SMEFT to the computation of EWPO, have a range of  applicability that goes well beyond the specific model analyzed here. They provide an important addition to any phenomenological analysis of a wide class of models featuring vector-like fermions, extended gauge groups, or TeV-scale implementations of the inverse-seesaw mechanism. 

Concerning the specific framework we have analysed,  we find that the new colored states of the model, while not
affecting EWPO at the tree level, generate a large flavor universal contribution at the loop level. This loop contribution comes via LL and NLL running, as well as finite one-loop matching contributions to the SMEFT operator $\cO_{HD}$. This effect dominates the overall improvement in the electroweak fit via an increase in the $W$-boson mass and universal shifts in several other $Z$-pole observables.
Furthermore, we observe good compatibility in the interplay of EWPO with other data, such as LFU tests at low energies and high-$p_T$ bounds, where the global best-fit point of the model gives a significant improvement over the SM hypothesis. In the case where the model addresses the charged-current $B$-anomalies, our combined analysis suggests the existence of new states not far above the TeV scale, giving effects in EWPO that cannot be decoupled. This sets a clear target for both current and near-future experiments, both at the intensity and high-energy frontiers.

\section*{Acknowledgements}

We thank Javier Fuentes Mart\'in and Felix Wilsch for useful discussions in relation to the one-loop matching of the model.
This project has received funding from the European Research Council (ERC) under the European Union's Horizon 2020 research and innovation programme under grant agreement 833280 (FLAY), and by the Swiss National Science Foundation (SNF) under contract 200020-204428.

\clearpage

\newpage
\appendix
    
\section{One-loop matching}\label{app:AppendixA}

\subsection{Box-diagram contributions to Higgs-fermion operators}

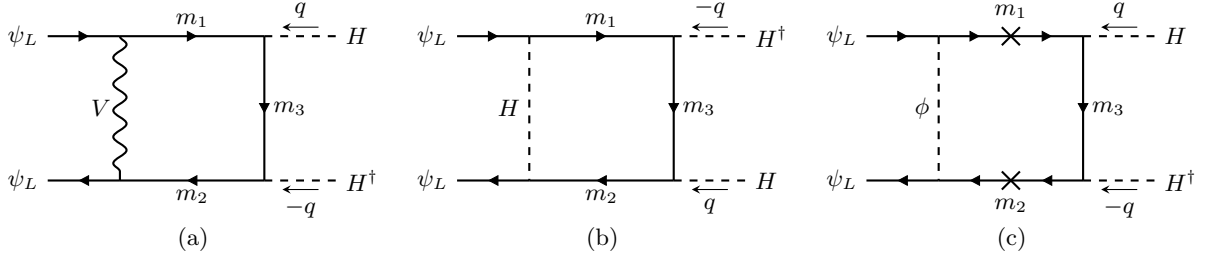
\begin{figure}[h!]
    \centering
    \begin{tabular}{c@{\hskip 0.3cm}c@{\hskip 0.3cm}c}
         \begin{tikzpicture}[thick,>=stealth,scale=1.2,baseline=-0.5ex]
            \draw[midarrow] (-1.6,0.8) node[left] {$\psi_L$} -- (-0.8,0.8);
            \draw[midarrow] (-0.8,0.8) -- (0.8,0.8);
            \draw[midarrow] (0.8,0.8) -- (0.8,-0.8);
            \draw[midarrow] (0.8,-0.8) -- (-0.8,-0.8);
            \draw[midarrow] (-0.8,-0.8) -- (-1.6,-0.8) node[left] {$\psi_L$};
            \draw[vector] (-0.8,0.8) -- (-0.8,-0.8);
            \draw[dashed] (0.8,0.8) -- (1.6,0.8) node[right] {$H$};
            \draw[dashed] (0.8,-0.8) -- (1.6,-0.8) node[right] {$H^\dagger$};
            \node[above] at (0,0.8) {$m_1$};
            \node[below] at (0,-0.8) {$m_2$};
            \node[right] at (0.8,0) {$m_3$};
            \node[left] at (-0.8,0) {$V$};
            \draw[thin,->] (1.4,0.9) -- (1,0.9);
            \node[above] at (1.2,0.9) {$q$};
            \draw[thin,->] (1.4,-0.9) -- (1,-0.9);
            \node[below] at (1.2,-0.9) {$-q$};
        \end{tikzpicture}
        &
        \begin{tikzpicture}[thick,>=stealth,scale=1.2,baseline=-0.5ex]
            \draw[midarrow] (-1.6,0.8) node[left] {$\psi_L$} -- (-0.8,0.8);
            \draw[midarrow] (-0.8,0.8) -- (0.8,0.8);
            \draw[midarrow] (0.8,0.8) -- (0.8,-0.8);
            \draw[midarrow] (0.8,-0.8) -- (-0.8,-0.8);
            \draw[midarrow] (-0.8,-0.8) -- (-1.6,-0.8) node[left] {$\psi_L$};
            \draw[dashed] (-0.8,0.8) -- (-0.8,-0.8);
            \draw[dashed] (0.8,0.8) -- (1.6,0.8) node[right] {$H^\dagger$};
            \draw[dashed] (0.8,-0.8) -- (1.6,-0.8) node[right] {$H$};
            \node[above] at (0,0.8) {$m_1$};
            \node[below] at (0,-0.8) {$m_2$};
            \node[right] at (0.8,0) {$m_3$};
            \node[left] at (-0.8,0) {$H$};
            \draw[thin,->] (1.4,0.9) -- (1,0.9);
            \node[above] at (1.2,0.9) {$-q$};
            \draw[thin,->] (1.4,-0.9) -- (1,-0.9);
            \node[below] at (1.2,-0.9) {$q$};
        \end{tikzpicture}
        &
        \begin{tikzpicture}[thick,>=stealth,scale=1.2,baseline=-0.5ex]
            \draw[midarrow] (-1.6,0.8) node[left] {$\psi_L$} -- (-0.8,0.8);
            \draw[midarrow] (-0.8,0.8) -- (0,0.8);
            \draw[] (0,0.8) +(-0.1,-0.1) -- +(0.1,0.1);
            \draw[] (0,0.8) +(-0.1,0.1) -- +(0.1,-0.1);
            \draw[midarrow] (0,0.8) -- (0.8,0.8);
            \draw[midarrow] (0.8,0.8) -- (0.8,-0.8);
            \draw[midarrow] (0.8,-0.8) -- (0,-0.8);
            \draw[] (0,-0.8) +(-0.1,-0.1) -- +(0.1,0.1);
            \draw[] (0,-0.8) +(-0.1,0.1) -- +(0.1,-0.1);
            \draw[midarrow] (0,-0.8) -- (-0.8,-0.8);
            \draw[midarrow] (-0.8,-0.8) -- (-1.6,-0.8) node[left] {$\psi_L$};
            \draw[dashed] (-0.8,0.8) -- (-0.8,-0.8);
            \draw[dashed] (0.8,0.8) -- (1.6,0.8) node[right] {$H$};
            \draw[dashed] (0.8,-0.8) -- (1.6,-0.8) node[right] {$H^\dagger$};
            \node[above] at (0,0.9) {$m_1$};
            \node[below] at (0,-0.9) {$m_2$};
            \node[right] at (0.8,0) {$m_3$};
            \node[left] at (-0.8,0) {$\phi$};
            \draw[thin,->] (1.4,0.9) -- (1,0.9);
            \node[above] at (1.2,0.9) {$q$};
            \draw[thin,->] (1.4,-0.9) -- (1,-0.9);
            \node[below] at (1.2,-0.9) {$-q$};
        \end{tikzpicture} \\
        (a) & (b) & (c)
    \end{tabular}
    \caption{Box diagrams contributing to the one-loop matching of the Higgs-current operators.}
    \label{fig:boxdiagrams}
\end{figure}
\noindent
The one-loop diagrams in the full theory proceed through box diagrams propagating fermions and either vector or scalar particles, as shown in~\cref{fig:boxdiagrams}. In dimensional regularisation with $d=4-2\epsilon$, the box diagram involving a heavy vector of mass $m_V = \sqrt{x_V} m_U$,~\cref{fig:boxdiagrams}(a), where $m_U$ is the $U_1$ leptoquark mass, reads (omitting the coupling structure and gauge group indices)
\begin{align}
    f_{\rm{VL}}(x_1,x_2,x_3,x_V)\,\bar{v}\slashed{q}{\rm{P}_{L}}u= \lim_{q^2\to 0}\int \frac{d^dk}{(2\pi)^d}\frac{\bar{v}\gamma_\mu \slashed{k}(\slashed{k}+\slashed{q})\slashed{k}\gamma^\mu{\rm{P}_{L}}u}{(k^2-x_1m_U^2)(k^2-x_2m_U^2)((k+q)^2-x_3m_U^2)(k^2-x_V m_U^2)}\,,
    \label{eq:fVL}
\end{align}
with $u\,,v$ being the external massless spinors whose chirality is projected with ${\rm{P}_{L}}=(1-\gamma_5)/2$, and we defined $m_i=\sqrt{x_i}m_U$, with $i=\{1,2,3\}$ as in~\cref{fig:boxdiagrams}. Similarly, in the case of the box diagram with a scalar exchange,~\cref{fig:boxdiagrams}(b,c), we distinguish two cases. The first case is when the scalar in question is the massless Higgs boson and the diagram reads
\begin{align}
f_{\rm{SL}}(x_1,x_2,x_3)\,\bar{v}\slashed{q}{\rm{P}_{L}}u= \lim_{q^2\to 0}\int \frac{d^dk}{(2\pi)^d}\frac{\bar{v}\slashed{k}(\slashed{k}-\slashed{q})\slashed{k}{\rm{P}_{L}}u}{(k^2-x_1m_U^2)(k^2-x_2m_U^2)((k-q)^2-x_3m_U^2)k^2}\,,
    \label{eq:fSL}
\end{align}
where the same notation, as in the case of a box diagram with the vector exchange, applies. The second case corresponds to the exchange of the $U_1$ Goldstone modes, and since we work in the Feynman gauge, the diagram reads
\begin{align}
    f_{\rm{SLR}}(x_1,x_2,x_3)\,\bar{v}\slashed{q}{\rm{P}_{L}}u= \sqrt{x_1\,x_2}\,m_U^2\lim_{q^2\to 0}\int \frac{d^dk}{(2\pi)^d}\frac{\bar{v}(\slashed{k}-\slashed{q}){\rm{P}_{L}}u}{(k^2-x_1m_U^2)(k^2-x_2m_U^2)((k-q)^2-x_3m_U^2)(k^2-m_U^2)}\,.
    \label{eq:fSLR}
\end{align}
In the Feynman gauge, the couplings of the $U_1$ Goldstone modes, $\phi_U$, to the fermions are~\cite{Fuentes-Martin:2020hvc}
\begin{align}
\label{eq:phiUcouplings}
\mathcal{L}_{\rm GB}&\supset\frac{g_4}{\sqrt{2}}\,\phi_U\left(\frac{m_L}{m_U}\,c_\ell\,W_{i2}\,\mathcal{\bar Q}_L^i\, L_R-\frac{m_Q}{m_U}\,c_q\,W_{2i}\,\bar Q_R\, \mathcal{L}_L^i -\frac{m_R}{m_U} \bar u_R^3 S_L\right)+{\rm h.c.},
\end{align}
where $\mathcal{Q}_L^1 = q_L^3$ and $\mathcal{Q}_L^2 = c_q Q_L - s_q q_L^2$, and similarly for leptons. In order to couple to the external Higgs bosons through couplings in~\cref{eq:LYukawa}, we need a chirality flip on the fermion lines corresponding to the masses $m_1$ and $m_2$ in~\cref{fig:boxdiagrams}(c), which reflects in the fact that the amplitude~\cref{eq:fSLR} is proportional to these masses. On the other hand, the couplings of the $G^\prime\,(Z^\prime)$ Goldstone bosons, $\phi_{G^\prime}\,(\phi_{Z^\prime})$ are
\begin{align}
\mathcal{L}_{\rm GB}\supset -i\frac{g_4}{2\sqrt{6}}\,\phi_{Z^\prime}\left(\frac{m_Q}{m_{Z^\prime}}\,s_q\,\bar q_L^\prime Q_R-3\frac{m_L}{m_{Z^\prime}}s_{\ell}\,\bar \ell_L^\prime L_R + 3 \,\frac{m_R}{m_{Z^\prime}}\,\bar S_L\, \nu^3_R\right)-ig_4\,\phi_{G^\prime}^a\frac{m_Q}{m_{G^\prime}}\,s_Q\,\bar q_L^\prime\, T^a Q_R+{\rm h.c.},
\end{align}
with $q_L^\prime = c_q q_L^2 + s_q Q_L$, and similarly for leptons. Notice that they do not allow for couplings to $q_L^3$ and $\ell_L^3$. As we are interested in the Higgs-fermion operators with at least one third-generation fermion, we neglect the effects of the $G^\prime$ and $Z^\prime$ Goldstone bosons.\\ In the following, we give explicit expressions for the loop functions that appear in the evaluation of Wilson coefficients, written in terms of $f_{\rm{VL}}$, $f_{\rm{SL}}$, and $f_{\rm{SLR}}$.
\begin{align}
    &  B_{\rm{VL}}(x_1,x_2,x_V) = -i\, 16 \pi ^2\, m_U^2\, f_{\rm{VL}}(0,x_1,x_2,x_V)\nonumber\\
    &= \frac{x_2}{\left(x_2-x_1\right) \left(x_2-x_V\right)}-\frac{x_V \left(x_2-2 x_V\right) \log \left(\frac{x_2}{x_V}\right)}{ \left(x_1-x_V\right) \left(x_2-x_V\right){}^2}+\frac{x_1 \left(2 x_1-x_2\right) \log \left(\frac{x_1}{x_2}\right)}{\left(x_1-x_2\right){}^2 \left(x_1-x_V\right)}\,,\\
    &\phantom{a}\nonumber\\
    & B_{\rm{SL}}(x_1,x_2) = -i\, 16 \pi ^2\, m_U^2\, f_{\rm{SL}}(x_1,x_1,x_2)\nonumber\\
    &= -\frac{x_1 }{ \left(x_1-x_2\right){}^2} + \frac{x_2 \left(x_2-3 x_1\right) \log \left(\frac{x_1}{x_2}\right)}{2 \left(x_2-x_1\right){}^3}\,,\\
    &\phantom{a}\nonumber\\
    & B_{\rm{V + S}}(x_1,x_2) = -i\, 16 \pi ^2\, m_U^2\, \big(f_{\rm{VL}}(x_1,x_1,x_2,1)+x_1 f_{\rm{SLR}}(x_1,x_1,x_2,1) \big)\nonumber\\
    &= -\frac{\left(4-2 x_2 x_1^2+x_1^2-2 x_2\right) \log \left(x_1\right)}{2 \left(x_1-1\right){}^2 \left(x_2-1\right){}^2}-\frac{x_1^3+2 x_2^2 x_1^2-7 x_2 x_1^2+4 x_1^2-2 x_2 x_1+2 x_2^2}{2 \left(x_1-1\right) \left(x_1-x_2\right){}^2 \left(x_2-1\right)}\nonumber\\
    &+\frac{x_2^2 \left(x_1^3-3 x_2 x_1^2+2 x_1^2+6 x_2 x_1-8 x_1-2 x_2^2+4 x_2\right) \log \left(\frac{x_1}{x_2}\right)}{2 \left(x_2-x_1\right){}^3 \left(x_2-1\right){}^2}\,.
\end{align}

\subsection{Triangle-diagram contributions to Higgs-quark operators}
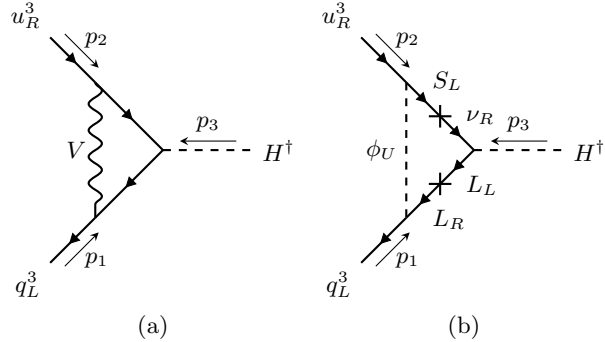
\begin{figure}[h]
    \centering  
    \begin{tabular}{cc}
    \begin{tikzpicture}[thick,>=stealth,scale=1.5]
        \draw[midarrow] (-1,1) node[above left] {$u_R^3$} -- (-0.6,0.6);
        \draw[midarrow] (-0.6,0.6) -- (0,0);
        \draw[midarrow] (0,0) -- (-0.6,-0.6);
        \draw[midarrow] (-0.6,-0.6) -- (-1,-1) node[below left] {$q_L^3$};
        \draw[vector] (-0.6,0.6) -- (-0.6,-0.6);
        \node[left] at (-0.6,0) {$V$};
        \draw[dashed] (0,0) -- (0.8,0) node[right] {$H^\dagger$};
        \begin{scope}[xshift=0.08cm,yshift=0.08cm]
            \draw[thin,->] (-0.95,0.95) -- (-0.65,0.65);
            \node[above right] at (-0.85,0.75) {$p_2$};
        \end{scope}
        \begin{scope}[xshift=0.08cm,yshift=-0.08cm]
            \draw[thin,->] (-0.95,-0.95) -- (-0.65,-0.65);
            \node[below right] at (-0.85,-0.75) {$p_1$};
        \end{scope}
        \begin{scope}[xshift=0.0cm,yshift=0.08cm]
            \draw[thin,->] (0.65,0) -- (0.15,0);
            \node[above] at (0.4,0) {$p_3$};
        \end{scope}
    \end{tikzpicture}
    &
    \begin{tikzpicture}[thick,>=stealth,scale=1.5]
        \draw[midarrow] (-1,1) node[above left] {$u_R^3$} -- (-0.6,0.6);
        \draw[midarrow] (-0.6,0.6) -- (-0.3,0.3);
        \node[above right] at (-0.45,0.45) {$S_L$};
        \draw[midarrow] (-0.3,0.3) -- (0,0);
        \node[above right] at (-0.15,0.15) {$\nu_R$};
        \draw[] (-0.3,0.3) +(0,0.1) -- +(0,-0.1);
        \draw[] (-0.3,0.3) +(0.1,0) -- +(-0.1,0);
        \draw[midarrow] (0,0) -- (-0.3,-0.3);
        \node[below right] at (-0.15,-0.15) {$L_L$}; 
        \draw[] (-0.3,-0.3) +(0,0.1) -- +(0,-0.1);
        \draw[] (-0.3,-0.3) +(0.1,0) -- +(-0.1,0);
        \draw[midarrow] (-0.3,-0.3) -- (-0.6,-0.6);
        \node[below right] at (-0.45,-0.45) {$L_R$};
        \draw[midarrow] (-0.6,-0.6) -- (-1,-1) node[below left] {$q_L^3$};
        \draw[dashed] (-0.6,0.6) -- (-0.6,-0.6);
        \node[left] at (-0.6,0) {$\phi_U$};
        \draw[dashed] (0,0) -- (0.8,0) node[right] {$H^\dagger$};
        \begin{scope}[xshift=0.08cm,yshift=0.08cm]
            \draw[thin,->] (-0.95,0.95) -- (-0.65,0.65);
            \node[above right] at (-0.85,0.75) {$p_2$};
        \end{scope}
        \begin{scope}[xshift=0.08cm,yshift=-0.08cm]
            \draw[thin,->] (-0.95,-0.95) -- (-0.65,-0.65);
            \node[below right] at (-0.85,-0.75) {$p_1$};
        \end{scope}
        \begin{scope}[xshift=0.0cm,yshift=0.08cm]
            \draw[thin,->] (0.65,0) -- (0.15,0);
            \node[above] at (0.4,0) {$p_3$};
        \end{scope}
    \end{tikzpicture}
    \\
    (a) & (b)
    \end{tabular}
    \caption{Triangle diagrams.}
    \label{fig:trianglediagrams1}
\end{figure}

\noindent
Another type of contribution to the Higgs-fermion operators comes from triangle diagrams as in~\cref{fig:trianglediagrams1}. 
At zero momentum, these renormalize the top Yukawa coupling, while the next order in the momenta gives, after applying the equations of motion (EOMs) for the massless external fermions, a contribution to $\cO_{Hq}^{(1,3)}$ operators.

\noindent Before applying EOMs, these amplitudes match onto three independent structures (operators in the Green's basis). Following the notation from \cite{Gherardi:2020det}, these are
\begin{align}
    Q_{uHD1} &= (\bar q_L u_R) D_\mu D^\mu \Tilde{H} \,, \\
    Q_{uHD3} &= (\bar q_L D_\mu D^\mu u_R) \Tilde{H} \,, \\
    Q_{uHD4} &= (\bar q_L D_\mu u_R) D^\mu \Tilde{H} \,.
\end{align}
The mapping of these onto the Warsaw basis operators reads
\begin{align}
    [\cC_{Hq}^{(1)}] = - [\cC_{Hq}^{(3)}] = \frac{1}{4} y_t {\rm Re} \, G_{uHD4} - \frac{1}{2} y_t {\rm Re} \, G_{uHD3} \,,
\end{align}
where $G_i$ indicate the Wilson coefficients in Green's basis and we have used $y_t \in \mathbb{R}$.
In order to compute the one-loop matching onto $G_{uHD3}$ and $G_{uHD4}$, one needs to choose three different momentum configurations and solve the linear system
\begin{align}
    A_{\rm EFT} \, \vec{G} = \vec{T}_{\rm UV} \,,
\end{align}
where $\vec{G}=(G_{uHD1},G_{uHD3},G_{uHD4})^\intercal$, the matrix $A_{\rm EFT}$ encodes the EFT amplitudes, and $\vec{T}_{\rm UV}$ is determined from the triangle diagrams for different momentum configurations. The final result for the Wilson coefficients is momentum independent, and we can make a convenient choice for the three configurations $(p_1,p_2,p_3)\in \{(q,0,-q),(q,-q,0),(0,q,-q)\}$, such that the result can be written as
\begin{align}
    G_{uHD1} &= \left. \partial_{q^2} F_T(q,0,-q) \right|_{q^2 = 0}\,,  \\
    G_{uHD3} &= \left. \partial_{q^2} F_T(q,-q,0) \right|_{q^2 = 0}\,, \\
    G_{uHD4} &= \left. \partial_{q^2} [ -F_T(0,q,-q) + F_T(q,-q,0) + F_T(q,0,-q) ] \right|_{q^2 = 0} \,,
\end{align}
where
\begin{align}
    \nonumber
   F_T (p_1,p_2,p_3) &= \frac{g_4^2}{24} y_t f_T(0,0,x_{Z'};p_1,p_2,p_3) + \frac{4}{3} g_4^2 y_t f_T(0,0,x_{G'};p_1,p_2,p_3) \\ \nonumber
   &+ \frac{g_4^2}{2} c_\chi y_\nu f_T(0,x_R,1;p_1,p_2,p_3)\\
   &+ \frac{g_4^2}{2} s_\chi Y_\nu^+\big( f_T(x_L,x_R,1;p_1,p_2,p_3)+ \sqrt{x_L\, x_R}\, f_{TG}(x_L,x_R,1;p_1,p_2,p_3)\big)
\end{align}
and
\begin{align}
    f_T(x_1,x_2,x_V;p_1,p_2,p_3) \bar{v} P_{\rm{R}} u&= \int \frac{d^dk}{(2\pi)^d}\frac{\bar{v}(p_1)\gamma_\mu(\slashed{k}+\slashed{p}_1)(\slashed{k}+\slashed{p}_1+\slashed{p}_3)\gamma^\mu P_{\rm{R}} u(p_2)}{((k+p_1)^2-x_1m_U^2)((k+p_1+p_3)^2-x_2m_U^2)(k^2-x_V m_U^2)}\,,
    \label{eq:fT} \\
    f_{TG}(x_1,x_2;p_1,p_2,p_3)\bar{v} P_{\rm{R}} u &= \int \frac{d^dk}{(2\pi)^d}\frac{\sqrt{x_1 x_2}\,m_U^2\bar{v}(p_1) P_{\rm{R}} u(p_2)}{((k+p_1)^2-x_1m_U^2)((k+p_1+p_3)^2-x_2m_U^2)(k^2- m_U^2)}\,.
    \label{eq:fTG} 
\end{align}

\noindent In the following, we give explicit expressions for the loop functions that appear in the evaluation of Wilson coefficients,
written in terms of $f_T$ and $f_{TG}$
\begin{align}
    & T_{\rm{V}}(x) = -i\, 16 \pi ^2\, m_U^2\, \left.\partial_{q^2}\bigg(f_{\rm{T}}(0,x,1;q,0,-q) - f_{\rm{T}}(0,x,1;0,q,-q)-f_{\rm{T}}(0,x,1;q,-q,0)\bigg)\right|_{q^2=0}\nonumber\\
    &= \frac{2 \left(1-3 x^2+2 x+(x+3) x \log (x)\right)}{(x-1)^3}\,,\\
    & T_{\rm{V + S}}(x_1,x_2) = -i\, 16 \pi ^2\, m_U^2\, \left.\partial_{q^2}\bigg(f_{\rm{TVG}}(x_1,x_2;q,0,-q) - f_{\rm{TVG}}(x_1,x_2;0,q,-q)-f_{\rm{TVG}}(x_1,x_2;q,-q,0)\bigg)\right|_{q^2=0}\nonumber\\
    &= -\frac{x_1^2 \left(2 x_2-x_1 \left(x_2-2\right)\right) \log \left(x_1\right)}{\left(x_1-1\right){}^2 \left(x_1-x_2\right){}^2}+\frac{x_2 \left(x_2+1\right) x_1^2-2 \left(3 x_2^2+x_2+1\right) x_1+2 x_2 \left(3 x_2+1\right)}{\left(x_1-1\right) \left(x_2-x_1\right) \left(x_2-1\right){}^2}\nonumber\\
    &+\frac{x_2^2 \left(2 x_1^2+\left(x_2-x_2^2-10\right) x_1+2 x_2 \left(x_2+3\right)\right) \log \left(x_2\right)}{\left(x_1-x_2\right){}^2 \left(x_2-1\right){}^3}\,,
\end{align}
where we define the function $f_{TVG}(x_1,x_2;p_1,p_2,p_3)$ as
\begin{equation}
    \label{eq:fTVG}
    f_{TVG}(x_1,x_2;p_1,p_2,p_3) = f_T(x_1,x_2,1;p_1,p_2,p_3) + \sqrt{x_1x_2}f_{TG}(x_1,x_2,1;p_1,p_2,p_3)\,.
\end{equation}

\subsection{Finite contribution to Higgs-lepton operators from inverse see-saw mechanism}
\begin{figure}[h!]
    \centering
        \begin{tabular}{cc}
            \begin{tikzpicture}[thick,>=stealth,scale=1.2,baseline=-0.5ex]
            \draw[] (0.0,-0.3) +(-0.1,-0.1) -- +(0.1,0.1);
            \draw[] (0.0,-0.3) +(-0.1,0.1) -- +(0.1,-0.1);
            \node[left] at (0.0,-0.3) {$m_R$};
            \draw[] (-0.5,0.9) +(-0.1,-0.1) -- +(0.1,0.1);
            \draw[] (-0.5,0.9) +(-0.1,0.1) -- +(0.1,-0.1);
            \draw[midarrow] (-1.5,0.9) node[left] {$\ell^3_L$} -- (-1,0.9);
            \draw[midarrow] (-1,0.9) -- (-0.5,0.9);
            \draw[midarrow] (-0.5,0.9) -- (0,0.9);
            \draw[midarrow] (0.0,0.9) -- (0.0,0.3);
            \draw[midarrow] (0.0,0.3) -- (0.0,-0.3);
            \draw[midarrow] (0.0,-0.3) -- (0.0,-0.9);
            \draw[midarrow] (0,-0.9) -- (-1.5,-0.9) node[left] {$\ell^3_L$};
            \draw[dashed] (0.0,0.3) -- (-1,0.9);
            \draw[dashed] (0.0,0.9) -- (0.9,0.9) node[right] {$H$};
            \draw[dashed] (0.0,-0.9) -- (0.9,-0.9) node[right] {$H^\dagger$};
            \node[left] at (-0.3,0.4) {$\phi_U$};
            \node[right] at (0,0) {$S_L$};
            \node[right] at (0,0.6) {$u_R^3$};
            \node[right] at (0,-0.6) {$\nu_R^3$};
            \node[above] at (-0.5,0.9) {$Q$};
            \draw[thin,<-] (0.3,1) -- (0.7,1);
            \node[above] at (0.5,1) {$q$};
            \draw[thin,<-] (0.3,-1) -- (0.7,-1);
            \node[below] at (0.5,-1) {$-q$};
        \end{tikzpicture} 
        &
        \begin{tikzpicture}[thick,>=stealth,scale=1.2,baseline=-0.5ex]
            \draw[] (0.0,0.3) +(-0.1,-0.1) -- +(0.1,0.1);
            \draw[] (0.0,0.3) +(-0.1,0.1) -- +(0.1,-0.1);
            \node[left] at (0.0,0.3) {$m_R$};
            \draw[] (-0.5,-0.9) +(-0.1,-0.1) -- +(0.1,0.1);
            \draw[] (-0.5,-0.9) +(-0.1,0.1) -- +(0.1,-0.1);
            \draw[midarrow] (-1.5,0.9) node[left] {$\ell^3_L$} -- (0,0.9);
            \draw[midarrow] (0.0,0.9) -- (0.0,0.3);
            \draw[midarrow] (0.0,0.3) -- (0.0,-0.3);
            \draw[midarrow] (0.0,-0.3) -- (0.0,-0.9);
            \draw[midarrow] (0,-0.9) -- (-0.5,-0.9);
            \draw[midarrow] (-0.5,-0.9) -- (-1,-0.9);
            \draw[midarrow] (-1,-0.9) -- (-1.5,-0.9) node[left] {$\ell^3_L$};
            \draw[dashed] (0.0,-0.3) -- (-1,-0.9);
            \draw[dashed] (0.0,0.9) -- (0.9,0.9) node[right] {$H$};
            \draw[dashed] (0.0,-0.9) -- (0.9,-0.9) node[right] {$H^\dagger$};
            \node[left] at (-0.3,-0.4) {$\phi_U$};
            \node[right] at (0,0) {$S_L$};
            \node[right] at (0,-0.6) {$u_R^3$};
            \node[right] at (0,0.6) {$\nu_R^3$};
            \node[below] at (-0.5,-0.9) {$Q$};
            \draw[thin,<-] (0.3,1) -- (0.7,1);
            \node[above] at (0.5,1) {$q$};
            \draw[thin,<-] (0.3,-1) -- (0.7,-1);
            \node[below] at (0.5,-1) {$-q$};
        \end{tikzpicture}
        \\
        (a) & (b) 
    \end{tabular}
    \caption{Diagrams contributing to the one-loop matching of the Higgs-lepton current operators when the inverse seesaw mechanism is implemented.}
    \label{fig:Sdiagram}
\end{figure}
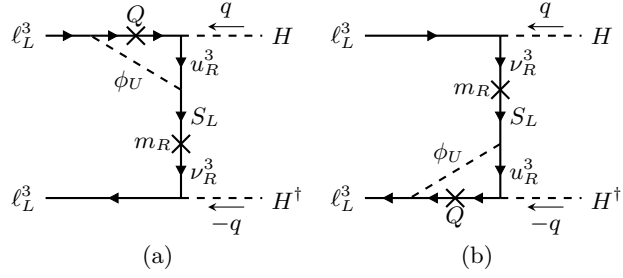

\noindent
The Yukawa couplings of the $U_1$ Goldstone boson $\phi_U$ in~\cref{eq:phiUcouplings} 
give contributions to the Higgs-lepton current operators $\cO_{H\ell}^{(1,3)}$ through the diagrams in~\cref{fig:Sdiagram}.

\noindent Expanding in $1/m_R$ and keeping only the leading term gives the contribution to dimension-6 operators.
Omitting Yukawa couplings, both diagrams give
\begin{equation}
\frac{i}{16\pi^2 m_U^2}f_{TS}(q^2)\,  \bar v \slashed{q} P_L u=\frac{m_Q}{m_R} \int \frac{d^dk}{(2\pi)^d}\frac{\bar{v} (\slashed{k}+\slashed{q})P_L u}{(k^2-m_Q^2)(k+q)^2(k^2-m_U^2)},
\end{equation}
with
\begin{align}
    f_{TS} (q^2 = 0) = \frac{\log x_Q}{2(1-x_Q)}  \,.
\end{align}
Notice that the Yukawa coupling involving $S_L$ in~\cref{eq:phiUcouplings} is proportional to $m_R$, so the dependence on $m_R$ will cancel.
Adding the two diagrams we get contributions to $[\mathcal{O}_{H\ell}^{(1)}]$ and $[\mathcal{O}_{H\ell}^{(3)}]$.

\subsection{Results: finite parts}
\noindent
In the following, we will present the results for the finite parts of the Wilson coefficients from the one-loop matching procedure.
In the previous sections, we have shown the contributions in the full UV theory. 
In order to obtain the matching, both tree-level and one-loop contributions in the SMEFT must be taken into account.

\begin{figure}[!h]
    \begin{tabular}{c@{\hskip 1cm}c@{\hskip 1cm}c@{\hskip 1cm}c}
        \begin{tikzpicture}[thick,>=stealth,baseline=-0.5ex]
            \draw[midarrow] (-1,1) -- (0,0);
            \draw[midarrow] (0,0) -- (-1,-1);
            \draw[dashed] (1,1) -- (0,0) -- (1,-1);
            \node[wc] at (0,0) {};
        \end{tikzpicture}
        &
        \begin{tikzpicture}[thick,>=stealth,baseline=-0.5ex]
         \draw[midarrow] (-0.8,0.8) -- (-0.05,0);
            \draw[midarrow] (-0.05,0) -- (-0.8,-0.8);
            \draw[midarrow] (0.8,-0.8) -- (0.05,0);
            \draw[midarrow] (0.05,0) -- (0.8,0.8);
            \draw[midarrow] (0.8,0.8) -- (0.8,-0.8);
            \draw[dashed] (0.8,0.8) -- (1.6,0.8);
            \draw[dashed] (0.8,-0.8) -- (1.6,-0.8);
        \end{tikzpicture}
        &
        \begin{tikzpicture}[thick,>=stealth,baseline=-0.5ex] 
            \draw[midarrow] (-0.8,0.8) -- (0,0.05);
            \draw[midarrow] (0,-0.05) -- (-0.8,-0.8);
            \draw[midarrow] (0.8,-0.8) -- (0,-0.05);
            \draw[midarrow] (0,0.05) -- (0.8,0.8);
            \draw[midarrow] (0.8,0.8) -- (0.8,-0.8);
            \draw[dashed] (0.8,0.8) -- (1.6,0.8);
            \draw[dashed] (0.8,-0.8) -- (1.6,-0.8);
        \end{tikzpicture}
        &
        \begin{tikzpicture}[thick,>=stealth,baseline=-0.5ex]
            \draw[midarrow] (-1.6,0.8) -- (-0.8,0.8);
            \draw[midarrow] (-0.8,0.8) -- (0,0);
            \draw[midarrow] (0,0) -- (-0.8,-0.8);
            \draw[midarrow] (-0.8,-0.8) -- (-1.6,-0.8);
            \draw[dashed] (0.8,-0.8) -- (0,0);
            \draw[dashed] (0,0) -- (0.8,0.8);
            \draw[dashed] (-0.8,0.8) -- (-0.8,-0.8);
            \node[wc] at (0,0) {};
        \end{tikzpicture}
        \\
        (a) & (b) & (c) & (d)
    \end{tabular}
    \caption{Diagrams in the SMEFT corresponding to box diagrams.}
    \label{fig:eftboxdiagrams}
\end{figure}
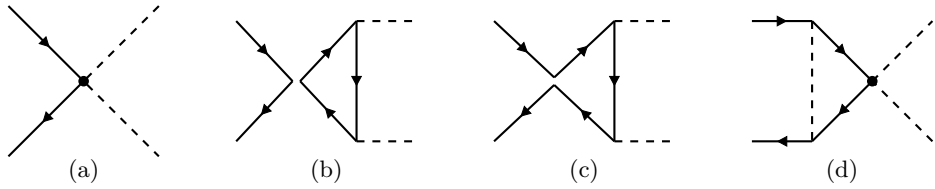

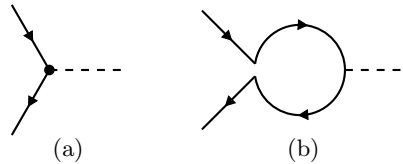
\begin{figure}[!h]
    \begin{tabular}{c@{\hskip 1cm}c@{\hskip 1cm}c@{\hskip 1cm}c}
        \begin{tikzpicture}[thick,>=stealth,baseline=-0.5ex]
            \draw[midarrow] (0,0) +(120:1) -- (0,0);
            \draw[midarrow] (0,0) -- +(-120:1);
            \draw[dashed] (0,0) -- (1,0);
            \node[wc] at (0,0) {};
        \end{tikzpicture}
        &
        \begin{tikzpicture}[thick,>=stealth,baseline=-0.5ex]
            \draw[midarrow] (172:0.6) +(135:1) -- (172:0.6);
            \draw[midarrow] (172:0.6) arc (172:0:0.6);
            \draw[midarrow] (0:0.6) arc (0:-172:0.6);
            \draw[midarrow] (-172:0.6) -- +(-135:1);
            \draw[dashed] (0:0.6) -- +(0:0.8);
        \end{tikzpicture}
        \\
        (a) & (b)
    \end{tabular}
    \caption{Diagrams in the SMEFT corresponding to triangle diagrams.}
    \label{fig:efttrianglediagrams}
\end{figure}

\noindent Diagrammatically, the UV amplitudes correspond to the ones in~\cref{fig:boxdiagrams} and~\cref{fig:trianglediagrams1}, while the SMEFT amplitudes to~\cref{fig:eftboxdiagrams} and~\cref{fig:efttrianglediagrams} for boxes and triangles, respectively. Equating the amplitudes in both theories, the finite pieces (FP) from matching read
\begin{align}
        16\pi^2\Lambda_U^2[\cC_{H\ell}^{(1)}]_{33}^{\rm{FP}} = &-\frac{3}{2}y_t^2c_\chi^2+  
        \frac{|y_\nu|^4}{g_4^2}B_{\rm{SL}}(x_R,0)+\frac{|y_\nu|^2|Y^\nu_+|^2}{g_4^2}B_{\rm{SL}}(x_R,x_L)\nonumber\\
        &-\frac{3}{2}y_t {\rm{Re}}(Y_+) s_\chi c_\chi B_{\rm{VL}}(x_Q,0,1) + \frac{3}{4}|Y_+|^2 s_\chi^2  B_{\rm{V+S}}(x_Q,0) + \frac{9}{48}|y_\nu|^2  B_{\rm{VL}}(0,x_R,x_{Z'})\nonumber\\
        &+\frac{1}{4} \, {\rm Re}\,(Y_+ y_{\nu}^{*})\, \frac{x_Q}{1-x_Q}\log\,x_Q\,,\\[3pt]
\label{CHl3FP}
        16\pi^2\Lambda_U^2[\cC_{H\ell}^{(3)}]_{33}^{\rm{FP}} = &\phantom{i} \frac{3}{2}y_t^2c_\chi^2+\frac{3}{2}y_t {\rm{Re}}(Y_+) s_\chi c_\chi B_{\rm{VL}}(x_Q,0,1) - \frac{3}{4}|Y_+|^2 s_\chi^2  B_{\rm{V+S}}(x_Q,0) - \frac{9}{48}|y_\nu|^2  B_{\rm{VL}}(0,x_R,x_{Z'})\nonumber\\
        &-\frac{1}{4} \, {\rm Re}\,(Y_+ y_{\nu}^{*})\, \frac{x_Q}{1-x_Q}\log\,x_Q\,,\\[3pt]
\label{CHq1FP}
 16\pi^2\Lambda_U^2[\cC_{Hq}^{(1)}]_{33}^{\rm{FP}} = &- \frac{1}{8} \frac{\Lambda_U^2}{m_Q^2} y_t^2 |Y_+|^2 + \frac{5 y_t^2}{6 x_{G'}}+ \frac{y_t^2}{24x_{Z'}}\nonumber\\
        & + \frac{1}{4}\bigg(|y_\nu|^2c_\chi^2 B_{\rm{VL}}(0,x_R,1) + {\rm{Re}}(y_\nu Y_+^{\nu*}) s_{2\chi} B_{\rm{VL}}(x_L,x_R,1)+ |Y_+^\nu|^2s_\chi^2 B_{\rm{V+S}}(x_L,x_R)\bigg)\nonumber\\
        & + \frac{1}{4}\bigg(y_t{\rm{Re}}( y_\nu) c_\chi T_{\rm{V}}(x_R)+ y_t{\rm{Re}}( Y_+^\nu) s_\chi T_{\rm{V+S}}(x_L,x_R)\bigg)\,,\\[3pt] 
\label{CHq3FP}
 16\pi^2\Lambda_U^2[\cC_{Hq}^{(3)}]_{33}^{\rm{FP}} = &\phantom{i} \frac{y_t^2}{6 x_{G'}}- \frac{y_t^2}{24x_{Z'}}\nonumber\\
        & - \frac{1}{4}\bigg(|y_\nu|^2c_\chi^2 B_{\rm{VL}}(0,x_R,1) + {\rm{Re}}(y_\nu Y_+^{\nu*}) s_{2\chi} B_{\rm{VL}}(x_L,x_R,1)+ |Y_+^\nu|^2s_\chi^2 B_{\rm{V+S}}(x_L,x_R)\bigg)\nonumber\\
        & - \frac{1}{4}\bigg(y_t{\rm{Re}}( y_\nu) c_\chi T_{\rm{V}}(x_R)+ y_t{\rm{Re}}( Y_+^\nu) s_\chi T_{\rm{V+S}}(x_L,x_R)\bigg)\,.
\end{align}

\subsection{Matching of $\cC_{HD}$}

\begin{figure}[!h]
\begin{tabular}{c@{\hskip 1cm}c@{\hskip 1cm}c}
\begin{tikzpicture}[thick,>=stealth,baseline=-0.5ex]
    \draw[dashed] (-1,1) -- (0,0) -- (-1,-1);
    \draw[dashed] (1,1) -- (0,0) -- (1,-1);
    \node[wc] at (0,0) {};
\end{tikzpicture}
&
\begin{tikzpicture}[thick,>=stealth,baseline=-0.5ex]
         \draw[dashed] (-0.05,0) -- (-0.8,0.8);
            \draw[dashed] (-0.05,0) -- (-0.8,-0.8);
            \draw[midarrow] (0.8,-0.8) -- (0.05,0);
            \draw[midarrow] (0.05,0) -- (0.8,0.8);
            \draw[midarrow] (0.8,0.8) -- (0.8,-0.8);
            \draw[dashed] (0.8,0.8) -- (1.6,0.8);
            \draw[dashed] (0.8,-0.8) -- (1.6,-0.8);
        \end{tikzpicture}
&
\begin{tikzpicture}[thick,>=stealth,baseline=-0.5ex]
    \draw[dashed] (-1.6,0.8) -- (-0.8,0.8);
    \draw[midarrow] (-0.8,0.8) -- (0.8,0.8);
    \draw[midarrow] (0.8,0.8) -- (0.8,-0.8);
    \draw[midarrow] (0.8,-0.8) -- (-0.8,-0.8);
    \draw[dashed] (-0.8,-0.8) -- (-1.6,-0.8);
    \draw[midarrow] (-0.8,-0.8) -- (-0.8,0.8);
    \draw[dashed] (0.8,0.8) -- (1.6,0.8);
    \draw[dashed] (0.8,-0.8) -- (1.6,-0.8);
\end{tikzpicture}
\\
(a) & (b) & (c)
\end{tabular}
\caption{Diagrams relevant for the matching to $C_{HD}$.}
\label{fig:CHDdiagrams}
\end{figure}
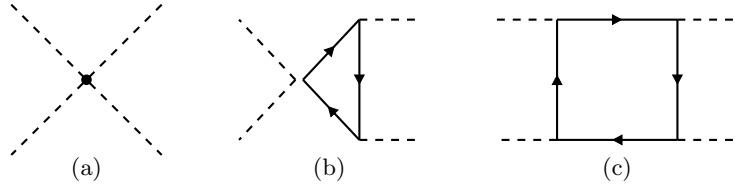
\noindent
The relevant diagrams for the matching of $\cC_{HD}$, both in the SMEFT and in the full 4321 model, are shown in~\cref{fig:CHDdiagrams}.
Similarly to the case of the triangle diagrams above, the four-point function receives contributions from four operators in the Green's basis \cite{Gherardi:2020det}:
\begin{align}
    {Q}_{H\Box} &= (H^\dagger H)\Box (H^\dagger H)\,,\\
     {Q}_{HD} &= |H^\dagger D_\mu H|^2\,,\\
     {Q}_{HD}^{\prime} &= (H^\dagger H)(D_\mu H)^\dagger(D^\mu H)\,,\\
     {Q}_{HD}^{\prime\prime} &= (H^\dagger H)D_\mu(i H^\dagger\overleftrightarrow{D}^\mu H)\,.
\end{align}
Therefore, we need to choose four different momentum configurations for the external Higgs states, $p_1\,,p_2\,,p_3$, and $p_4$, set up the system for $G_{H\Box}\,,G_{HD}\,,G_{HD}^{\prime}$, and $G_{HD}^{\prime\prime}$, and solve for $G_{HD}=\cC_{HD}$. We obtain
\begin{align}
        16\pi^2[\cC_{HD}]^{\rm{FP}} &= \frac{|Y_+|^2}{m_Q^2}\bigg(9y_t^2-2|Y_+|^2\bigg)-\frac{|y_\nu|^4}{2m_R^2}+ \frac{|Y_+^\nu|^4}{m_R^2}B^\ell_1(x_{LR})+\frac{|y_\nu|^2|Y_+^\nu|^2}{m_R^2}B^\ell_2(x_{LR})\,,
\end{align}
with the loop functions being
\begin{align}
    B^\ell_1(x) &= \frac{6 (5-x) x^2 \log (x)+(23-x (x (15-4 x)+9)) x-3}{6 (1-x)^5}\,,\\[4pt]
    B^\ell_2(x) &= \frac{2 x^2 \log (x)-3 x^2+4 x-1}{(1-x)^3}\,,
 \end{align}
 and $x_{LR} = x_L/x_R$.

\subsection{Semileptonic Wilson coefficients at one loop}\label{app:RWC@1loop}
\noindent
The relevant four-fermion Wilson coefficients for $b\to c\tau \nu$ transitions and high-$p_T$, improved with the NLO corrections in $SU(4)_h$ gauge coupling, where $\alpha_4=g_4^2/4\pi$, read
\begin{align}
    [\cC_{\ell q}^{(1)}]_{3333} &= -\frac{c_\chi^2}{2\LU^2}\left[1+\frac{\alpha_4}{4\pi}\left(\frac{17}{12}f_1(x_{Z^\prime})+\frac{4}{3}f_1(x_{G^\prime})\right)\right]
    +\frac{1}{4\Lambda_U^2\xZ}\left[1 +\frac{\alpha_4}{4\pi}\frac{3}{8}\right]\nonumber\\
    &\,\,+\frac{2}{\LU^2}\frac{\alpha_4}{4\pi}\bigg(c_\chi^4 + 2s_\chi^2 c_\chi^2\, f_3(x_Q,x_L)+s_\chi^4\, f_4(x_Q,x_L)\bigg)\,,\\
    [\cC_{\ell q}^{(3)}]_{3333} &= -\frac{c_\chi^2}{2\LU^2}\left[1+\frac{\alpha_4}{4\pi}\left(\frac{17}{12}f_1(x_{Z^\prime})+\frac{4}{3}f_1(x_{G^\prime})\right)\right]\,,\\
    [\cC_{\ell e d q}]_{3333} &= \frac{2 c_\chi e^{i\phi_R}}{\LU^2}\left[1+\frac{\alpha_4}{4\pi}\left(\frac{23}{12}f_1(x_{Z^\prime})+\frac{16}{3}f_1(x_{G^\prime})\right)\right]\,,\\
     [\cC_{\ell q}^{(3)}]_{3323} &= -\frac{s_q s_\chi c_\chi}{2\LU^2}\left[1+\frac{\alpha_4}{4\pi}\bigg(\frac{7}{8}f_1(x_{Z^\prime})+\frac{13}{24}\,f_2(x_{Z^\prime},x_Q)+\frac{4}{3}\,f_2(x_{G^\prime},x_Q)\,\bigg)\right]\,,\\
    [\cC_{\ell e d q}]_{3332} &= \frac{2 s_q s_\chi e^{i\phi_R}}{\LU^2}\left[1+\frac{\alpha_4}{4\pi}\bigg(\frac{13}{8}f_1(x_{Z^\prime})+\frac{7}{24}\,f_2(x_{Z^\prime},x_Q)+\frac{16}{3}\,f_2(x_{G^\prime},x_Q)\bigg)\right]\,,\\
    [\cC_{e d}]_{3333} &= -\frac{1}{\LU^2}\left[1+\frac{\alpha_4}{4\pi}\left(\frac{17}{12}f_1(\xZ)+\frac{4}{3}f_1(\xG)\right)\right]+\frac{1}{4 \LU^2\xZ}\left[1+\frac{\alpha_4}{4\pi}\frac{3}{8}\right]+\frac{2}{\LU^2}\frac{\alpha_4}{4\pi}\,,\\
    [\cC_{\ell d}]_{3333} &= \frac{1}{4\Lambda_U^2\xZ}\left[1-\frac{\alpha_4}{4\pi}\frac{3}{8}\right] + \frac{1}{2\LU^2}\frac{\alpha_4}{4\pi}\left(c_\chi^2+s_\chi^2 f_5(x_L)\right)
 \,,\\
   [\cC_{qe}]_{3333} &= \frac{1}{4\Lambda_U^2\xZ}\left[1-\frac{\alpha_4}{4\pi}\frac{3}{8}\right] + \frac{1}{2\LU^2}\frac{\alpha_4}{4\pi}\left(c_\chi^2+s_\chi^2 f_5(x_Q)\right)
 \,,
\end{align}
where the loop functions are
\begin{align}
    f_1(x_V) &= \frac{\log(x_V)}{x_V-1}\,,\\
    f_2(x_V,x_Q) &=\frac{1}{x_V-x_Q}\left(\frac{x_V\log(x_V)}{x_V-1}- \frac{x_Q\log(x_Q)}{x_Q-1} \right)\,, \\
    f_3(x_Q,x_L) &= \frac{1}{2}\left(\frac{1}{1-x_Q}+\frac{1}{1-x_L}+ \frac{x_Q\log(x_Q)}{(1-x_Q)^2}+\frac{x_L\log(x_L)}{(1-x_L)^2}\right)\,,\\
    f_4(x_Q,x_L) &= \frac{1}{(1-x_Q)(1-x_L)}+\frac{x_Q^2\log(x_Q)}{(1-x_Q)^2(x_Q-x_L)}-\frac{x_L^2\log(x_L)}{(1-x_L)^2(x_Q-x_L)}\nonumber\\
    &-\frac{7x_Q x_L}{16(1-x_Q)(1-x_L)}+\frac{(8-x_Q)x_Lx_Q^2\log(x_Q)}{16(1-x_Q)^2(x_Q-x_L)}+\frac{(8-x_L)x_Qx_L^2\log(x_L)}{16(1-x_L)^2(x_Q-x_L)}\,,\\
    f_5(x) &= \frac{1-x+x\log(x)}{(1-x)^2}\,.
\end{align}
In the limit of vanishing vector-like fermion masses, $x_{Q,L}\to 0$, the loop functions $f_{i=3,4,5}$ go to 1, while $f_2(x_V,0) = f_1(x_V)$, thus reproducing the results of Ref.~\cite{Fuentes-Martin:2019ign,Fuentes-Martin:2020luw}.

\section{Running effects}
\label{app:AppendixB}

We include the running of the Wilson coefficients from the UV to the EW scale. This running can be found solving the RGE equations:
\begin{equation}
\mu\frac{d}{ \mu} \cC (\mu) =  \mathcal{A}(\mu) \,\cC(\mu),
\end{equation}
where $\cC$ is a vector with all dimension 6 Wilson coefficients, including flavor indices, and $\mathcal{A}$ is the anomalous dimension matrix, that can be read from~\cite{{Jenkins:2013zja,Jenkins:2013wua,Alonso:2013hga}}:
\begin{equation}
\mathcal{A}(\mu)=
\frac{y_t(\mu)^2}{16\pi^2} \mathcal{A}_t+
\frac{g_s(\mu)^2}{16\pi^2}  \mathcal{A}_s+
\frac{g_L(\mu)^2}{16\pi^2} \mathcal{A}_L + \ldots
\end{equation}
The anomalous dimension matrix depends on $\mu$ through the running of the SM parameters. Notice that the impact of dimension 6 operators on the SM parameters running is a dimension 8 effect~\cite{Fuentes-Martin:2020zaz} so it can be neglected. The solution to this differential equation is
\begin{align}
\cC(\mu) &= \mathcal{P} \int_{\mu_{0}}^{\mu}
 {\rm exp}\,\mathcal{A}(\mu)
\,d\log \mu \,\cC(\mu_{0})\nonumber\\
&=\left(\mathbb{1} +\int_{\mu_{0}}^{\mu}  d\log \mu\, \mathcal{A}(\mu) +
\int_{\mu_{0}}^{\mu}d\log \mu_1 \int_{\mu_{0}}^{\mu_1} d\log \mu_2 
\mathcal{A}(\mu_1)\mathcal{A}(\mu_2)
+\ldots
\right)\cC(\mu_{0}),\label{eq:SolutionRun}
\end{align}
where $\mathcal{P}$ is the $\mu$-ordered product.
If we neglect the running of the SM parameters,~\cref{eq:SolutionRun} reads
\begin{align}
\cC(\mu) =\left( \mathbb{1}+\log\left( \frac{\mu}{\mu_{0}}\right)\mathcal{A}+
\frac{1}{2}\log^2 \left(\frac{\mu}{\mu_{0}}\right)
\mathcal{A}^2 + \ldots \right)
\cC(\mu_{0}).\label{eq:SolutionRunNoSmR}
\end{align}

\subsection{Leading-log running in $y_t$}
\noindent
The one-integral term of the expansion in~\cref{eq:SolutionRun} corresponds to the leading-log running. In this work, the most important leading-log contribution to the operators relevant for the EW fit is the one induced by the top Yukawa ($g_s$ does not induce any leading-log running in the sector relevant for our model):
\begin{align}
\cC(\mu)-\cC(\mu_0)=&
\frac{1}{16\pi^2}\mathcal{A}_t \cC(\mu_0)\int_{\mu_0}^{\mu} y_t(\mu)^2 d \log \mu \nonumber \\
=&
\frac{\bar y_t^2}{16\pi^2}\mathcal{A}_t \cC(\mu_0)\log \frac{\mu}{\mu_0} \,,
\end{align}
where $y_t^2$ is the average of $y^2_t(\mu)$,
\begin{equation}
\bar y_t^2= \frac{1}{\log\frac{\mu}{\mu_0}}\int_{\mu_{0}}^{\mu}\, d \log \mu^{\prime} \,y_t^2(\mu^{\prime}).\label{eq:logaverageyt}
\end{equation}
Substituting $\mathcal{A}_t$, we recover the leading-log formulas given in section~\ref{sec:LL}. Running from $2.5\,$TeV to $m_t$, and using DsixTools~\cite{Fuentes-Martin:2020zaz}, we get $\bar y_t \approx 0.87$, which is the value we use for $y_t$ in this work.

\subsection{Next-to-leading-log running in $y_t$ and $g_s$}
\label{app:AppendixBNLL}
\noindent
The two-integral term of~\eqref{eq:SolutionRun}, or the $\log^2$ term of~\eqref{eq:SolutionRunNoSmR}, give the next-to-leading-log running.
Including $y_t$ and $g_s$ effects, using the tree-level expressions of the third-family Wilson coefficients of~\cref{tab:fourfermionoperators}, and running from the UV, we get:

\begin{align}
\label{eq:CHDNLLAp}[\cC_{HD}]^{\rm NLL}=&-\frac{3y_t^4}{32\pi^4 \Lambda^2_Ux_{G^{\prime}}}\log^2 \left(\frac{\mu^2}{m_{G^{\prime}}^2}\right) +\frac{81 y_t^4 |Y_+|^2}{512 \pi^4 m_Q^2}
\log ^2\left(\frac{\mu^2}{m_Q^2}\right)-
\frac{3y_t^4}{1024\pi^4 \Lambda^2_Ux_{Z^{\prime}}}\log^2 \left(\frac{\mu^2}{m_{Z^{\prime}}^2}\right)
\,,\\
[\cC^{(1)}_{Hq}]_{33}^{\rm NLL}=&-\frac{242g_s^2 y_t^2+153 y_t^4}{6912\,\pi^4 \Lambda^2_Ux_{G^{\prime}}}\log^2 \left(\frac{\mu^2}{m_{G^{\prime}}^2}\right)
+\frac{45 y_t^4 |Y_+|^2}{4096 \pi^4 m_Q^2}
\log ^2\left(\frac{\mu^2}{m_Q^2}\right)
-
\frac{176 \, g_s^2 y_t^2+153\,y_t^4}{221184\,\pi^4 \Lambda^2_Ux_{Z^{\prime}}}\log^2 \left(\frac{\mu^2}{m_{Z^{\prime}}^2}\right)
\,,\\
[\cC^{(3)}_{Hq}]_{33}^{\rm NLL}=&\frac{26\,g_s^2 y_t^2+99\, y_t^4}{6912\, \pi^4 \Lambda^2_Ux_{G^{\prime}}}\log^2 \left(\frac{\mu^2}{m_{G^{\prime}}^2}\right)
-\frac{3 y_t^4 |Y_+|^2}{2048 \pi^4 m_Q^2}
\log ^2\left(\frac{\mu^2}{m_Q^2}\right)
+
\frac{176 \, g_s^2 y_t^2+99\,y_t^4}{221184\,\pi^4 \Lambda^2_Ux_{Z^{\prime}}}\log^2 \left(\frac{\mu^2}{m_{Z^{\prime}}^2}\right)
\,,\\
[\cC^{(1)}_{Hl}]_{33}^{\rm NLL}=&  - \frac{27}{21} [\cC^{(3)}_{Hl}]_{33}^{\rm NLL}= -\frac{27 y_t^4  c_\chi^2}{2048\,\pi^4 \Lambda^2_U}\log^2 \left(\frac{\mu^2}{m_{U}^2}\right)
+\frac{27 y_t^4 |y_{\nu}|^2}{4096 \pi^4 m_R^2} \log^2 \left( 
\frac{\mu^2}{m_R^2}
\right)
\,,\\
[\cC_{Hd}]_{33}^{\rm NLL}=& \frac{y_t^2 g_s^2}{32\pi^4 \Lambda^2_Ux_{G^{\prime}}}\log^2 \left(\frac{\mu^2}{m_{G^{\prime}}^2}\right)\, \label{eq:NLLbR}
.
\end{align}
These expressions include the running of the SM parameters if for $y_t$ we use the average $\bar y_t$ of~\cref{eq:logaverageyt}, and for $g_s$, the average
\begin{equation}
\bar g^2_s=\frac{2}{\bar y_t^2 \log^2 \frac{\mu}{\mu_0}} \int_{\mu_{0}}^{\mu}d\log \mu_1 \int_{\mu_{0}}^{\mu_1} d\log \mu_2 \,
y_t^2(\mu_1)\,g_s^2(\mu_2).
\end{equation}
Running from $2.5\,$TeV to $m_t$, and using DsixTools~\cite{Fuentes-Martin:2020zaz}, we get $\bar g_s \approx 1.1$.

\subsection{Leading-log running in $g_{L}$}
\label{sec:gLrunning}
\noindent
The weak running in $g_L$ gives contributions to $\cC_{H\ell}^{(3)}$ and $\cC_{Hq}^{(3)}$, but not to the corresponding singlet Wilson coefficients. Taking the leading log, neglecting the running of $g_L$, and using that $[\cC_{\ell\ell}]_{ijkl}$ is real, we find that the third-family contributions are
\begin{align}
    \frac{16\pi^2}{g_{L}^2} [\cC_{H\ell}^{(3)}]_{ij}^{\rm{LL}} &= \frac{N_c}{3} [\cC_{\ell q}^{(3)}]_{ij 33} \log\bigg(\frac{\mu^2}{m_U^2}\bigg)  - \bigg(\frac{17}{6}[\cC_{H\ell}^{(3)}]_{ij} -\frac{1}{3}[\cC_{H\ell}^{(3)}]_{33}\delta_{ij} \bigg)\log\bigg(\frac{\mu^2}{m_R^2}\bigg) + \frac{1}{3}[\cC_{\ell\ell}]_{i 33j}\log\bigg(\frac{\mu^2}{m_{Z'}^2}\bigg)\,.
\end{align}
Similarly, using that $[\cC_{qq}^{(1,3)}]_{ijkl}$ is real, we find
\begin{align}
    \frac{16\pi^2}{g_{L}^2} [\cC_{Hq}^{(3)}]_{ij}^{\rm{LL}} &=\frac{1}{3} [\cC_{\ell q}^{(3)}]_{33ij} \log\bigg(\frac{\mu^2}{m_U^2}\bigg) +\frac{1}{3} [\cC_{H\ell}^{(3)}]_{33}\, \delta_{ij} \log\bigg(\frac{\mu^2}{m_R^2}\bigg) \nonumber \\
    &+\frac{2}{3} N_c [\cC_{qq}^{(3)}]_{ij33}\log\bigg(\frac{\mu^2}{m_{G'}^2}\bigg)   +\frac{1}{3} \sum_V \bigg( [\cC_{qq}^{(1)}]^{V}_{i33j} - [\cC_{qq}^{(3)}]^{V}_{i33j} \bigg) \log\bigg(\frac{\mu^2}{m_{V}^2}\bigg) \,, 
\end{align}
where the sum on $V=G',Z'$ is understood to be over the $G'$ and $Z'$ contributions to $\cC_{qq}^{(1,3)}$. In terms of the model parameters, the Wilson coefficients read
\begin{align}
[\cC_{H\ell}^{(3)}]_{ij}^{\rm LL} =& -\frac{g_L^2}{32\pi^2} \left[\frac{1}{\LU^2}\left(c_\chi^2\log\bigg(\frac{\mu^2}{m_U^2}\bigg)+\frac{1}{4\xZ}\log\bigg(\frac{\mu^2}{m_{Z'}^2}\bigg)\right)  - \frac{17|y_{\nu}|^2}{12m_{R}^2} \log\bigg(\frac{\mu^2}{m_{R}^2}\bigg)\right]\delta_{i 3}\delta_{j 3} -\frac{g_L^2 |y_{\nu}|^2}{192\pi^2 m_{R}^2} \log\bigg(\frac{\mu^2}{m_{R}^2}\bigg)\delta_{ij} \,,
\end{align}
and similarly
\begin{align}
[\cC_{Hq}^{(3)}]_{ij}^{\rm LL} = -\frac{g_L^2}{12\pi^2\LU^2}\left[\frac{c_\chi^2}{8} \log\bigg(\frac{\mu^2}{m_U^2}\bigg)+\frac{1}{3\xG} \log\bigg(\frac{\mu^2}{m_{G'}^2}\bigg) + \frac{1}{96\xZ} \log\bigg(\frac{\mu^2}{m_{Z'}^2}\bigg)\right]\delta_{i3} \delta_{j3}
-\frac{g_L^2 |y_{\nu}|^2}{192\pi^2 m_{R}^2} \log\bigg(\frac{\mu^2}{m_{R}^2}\bigg)\delta_{ij}
&
\,.
\end{align}

\section{Electroweak observables at 1-loop}\label{app:EWPO}

\subsection{Connection with SMEFT}
\label{sec:SMEFTdelta}
\noindent
Inserting the generated SMEFT operators induces vertex and mass modifications of the EW gauge bosons. Working in the $\{\alpha_{EM},m_Z,G_F\}$ input scheme, the relevant terms of the effective Lagrangian for the EW fit become~\cite{Breso-Pla:2021qoe}
\begin{align}
\mathcal{L}_{\rm eff} \supset &- \frac{g_L}{\sqrt{2}}W^{+\mu}\left[\bar u_L^i \gamma_{\mu} \left(V_{ij}+\delta g_{ij}^{Wq}\right)d_L^j + \bar \nu_L^i \gamma_{\mu} \left(\delta_{ij}+\delta g_{ij}^{W \ell }\right)e_L^j\right]+{\rm h.c.}\nonumber \\
&-\sqrt{g_L^2+g_Y^2}\, Z^{\mu} \left( \bar f^i_L \gamma_{\mu}
\left( g_L^{Zf}\delta_{ij}+ \delta g_{L\,ij}^{Zf}\right) f_L^j
+\bar f^i_R \gamma_{\mu}
\left[ g_R^{Zf} \delta_{ij}+ \delta g_{R\,ij}^{Zf}\right] f_R^j
\right)\nonumber \\
&+\frac{g_L^2 v^2}{4}(1+\delta m_W)^2 W^{+\mu} W^-_{\mu}+\frac{g_L^2 v^2}{8 c_W^2}Z^{\mu} Z_{\mu},
\label{EWEffLag}
\end{align}
where
\begin{align}
g_L^{Zf}=\,T_f^3-s_W^2 Q_f ,~~~~~~g_R^{Zf}= -s_W^2 Q_f .
\end{align}
The vertex modifications of the $Z$-boson are
\begin{align}
\delta g_{L\,ij}^{Z\nu}=&-\frac{v^2}{2}\left([\cC_{Hl}^{(1)}]_{ij}-[\cC_{Hl}^{(3)}]_{ij}\right)+\delta^U(1/2,0)\, \delta_{ij}+[\delta g_{L\,ij}^{Z\nu}]_{\rm loop},\\
\delta g_{L\,ij}^{Ze}=&-\frac{v^2}{2}\left([\cC_{Hl}^{(1)}]_{ij}+[\cC_{Hl}^{(3)}]_{ij}\right)+\delta^U(-1/2,-1)\delta_{ij}+[\delta g_{L\,ij}^{Ze}]_{\rm loop}\\
\delta g_{R\,ij}^{Ze}=&\,-\frac{v^2}{2}[\cC_{He}]_{ij}+\delta^U(0,-1)\, \delta_{ij}+[\delta g_{R\,ij}^{Ze}]_{\rm loop},\\
\delta g_{L\,ij}^{Zu}=&\,-\frac{v^2}{2}V_{ik} \left(
[\cC_{Hq}^{(1)}]_{kl}-[\cC_{Hq}^{(3)}]_{kl}
\right) V_{lj}^{\dagger}+\delta^U(1/2,2/3)\, \delta_{ij}+[\delta g_{L\,ij}^{Zu}]_{\rm loop},\\
\delta g_{R\,ij}^{Zu}=&\,-\frac{v^2}{2}[\cC_{Hu}]_{ij}+\delta^U(0,2/3)\, \delta_{ij}+[\delta g_{R\,ij}^{Zu}]_{\rm loop}
,\\
\delta g_{L\,ij}^{Zd}=&-\frac{v^2}{2}\left([\cC_{Hq}^{(1)}]_{ij}+[\cC_{Hq}^{(3)}]_{ij}\right)+\delta^U(-1/2,-1/3)\, \delta_{ij}+[\delta g_{L\,ij}^{Zd}]_{\rm loop},\label{eq:ZtobLbL}\\
\delta g_{R\, ij}^{Zd}=&\,-\frac{v^2}{2}[\cC_{Hd}]_{ij}+\delta^U(0,-1/3)\, \delta_{ij}+[\delta g_{R\, ij}^{Zd}]_{\rm loop},
\end{align}
where $\delta^U(T^3,Q)$ is a family-universal contribution that is given by
\begin{align}
\delta^U(T^3,Q)=&
-v^2\left( T^3+Q\frac{g_Y^2}{g_L^2-g_Y^2} \right)
\left(\frac{1}{4}\cC_{HD}+\frac{1}{2}[\cC^{(3)}_{H\ell}]_{22}+\frac{1}{2}[\cC^{(3)}_{H\ell}]_{11}-\frac{1}{4}[\cC_{\ell\ell}]_{1221}\right)
\nonumber\\
&-v^2Q\frac{g_Lg_Y}{g_L^2-g_Y^2} \cC_{HWB}
+\delta^U_{\rm loop}(T^3,Q).
\end{align}

\begin{figure}
    \centering
    \begin{tabular}{c@{\hskip 1cm}c@{\hskip 1cm}c}
        \begin{tikzpicture}[thick,>=stealth,baseline=-0.5ex]
            \draw[vector] (-1.2,0) -- (0,0);
            \draw[midarrow] (0,0) -- (45:1);
            \draw[midarrow] (45:1) -- (45:1.5);
            \draw[midarrow] (-45:1.5) -- (-45:1);
            \draw[midarrow] (-45:1) -- (-0,0);
            \draw[dashed] (45:1) -- (-45:1);
            \node[wc] at (0,0) {};
            \node[left] at (-1.2,0) {$Z$};
            \node[above] at (0.3,0.35) {$t$};
            \node[below] at (0.3,-0.35) {$t$};
            \node[right] at (45:1.5) {$b_L$};
            \node[right] at (-45:1.5) {$b_L$};
            \node[right] at (0.7,0) {$H^{\pm}$};
            \node[above] at (-0.25,0.05) {$\delta g_{R33}^{Zu}$};
        \end{tikzpicture}
        &
        \begin{tikzpicture}[thick,>=stealth,baseline=-0.5ex]
            \draw[vector] (-1.7,0) -- (-0.7,0);
            \draw[midarrow] (-0.7,0) arc (180:0:0.7);
            \draw[midarrow] (0.7,0) arc (0:-180:0.7);
            \draw[midarrow] (0.7,0) -- +(45:1.2);
            \draw[midarrow] (0.7,0) +(-45:1.2) -- (0.7,0);
            \node[wc] at (0.7,0) {};
            \node[left] at (-1.7,0) {$Z,W$};
            \node[above] at (0,0.7) {$t$};
            \node[below] at (0,-0.7) {$t,b$};
        \end{tikzpicture}
        &
        \begin{tikzpicture}[thick,>=stealth,baseline=-0.5ex]
            \draw[vector] (-1.7,0) -- (-0.7,0);
            \draw[midarrow] (-0.7,0) arc (180:0:0.7);
            \draw[midarrow] (0.7,0) arc (0:-180:0.7);
            \draw[vector] (0.7,0) -- (1.7,0);
            \node[wc] at (-0.7,0) {};
            \node[left] at (-1.7,0) {$Z$};
            \node[right] at (1.7,0) {$Z$};
            \node[above] at (0,0.7) {$t$};
            \node[below] at (0,-0.7) {$t$};
            \node[right] at (-0.7,0.0) {$\delta g_{R33}^{Zu}$};
        \end{tikzpicture}
        \\
        (a) & (b) & (c)
    \end{tabular}
    \caption{Diagrams contributing to the electroweak observables at one-loop.}
    \label{fig:EW1loopdiagrams}
\end{figure}
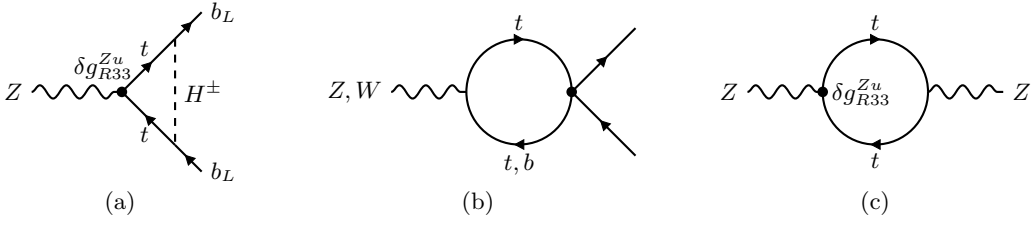

\noindent
To be consistent with the one-loop matching, we include one-loop corrections in $y_t$. In our model, they are given by the diagrams in~\cref{fig:EW1loopdiagrams}. Notice that neglecting terms $O(s_{q,\ell})$ and $O(g^2_{s,Y}/g_4^2)$, diagrams (a) and (b) only affect the purely third-family vertices.
In principle, diagram (c) affects the $Z$-mass, but due to the $\{\alpha_{EM},m_Z,G_F\}$ input scheme we are using this correction is translated into a correction of the flavor-universal shift $\delta^U$ and the $W$-mass we show below.
The one-loop corrections in $y_t$ to the $Z$-vertex modifications relevant for the EW fit (which excludes the $Z\bar t t$ coupling) in our model are
\begin{align}
\delta^U_{\rm loop}(T^3,Q)=&
-\left( T^3+Q\frac{g_Y^2}{g_L^2-g_Y^2} \right)
 \frac{3 m_t^2}{8 \pi^2} [\cC_{Hu}]_{33} \log\frac{\mu_{\rm EW}^2}{m_t^2},\\
[\delta g_{L\,ij}^{Z\nu}]_{\rm loop}=&\frac{3 m_t^2}{16 \pi^2} \left([\cC_{\ell q}^{(1)}]_{3333}+[\cC_{\ell q}^{(3)}]_{3333}-[\cC_{\ell u}]_{3333}\right)\delta_{i3}\delta_{j3}\log \frac{\mu_{\rm EW}^2}{m_t^2},\\
[\delta g_{L\,ij}^{Ze}]_{\rm loop}=&\frac{3 m_t^2}{16 \pi^2} \left([\cC_{\ell q}^{(1)}]_{3333}-[\cC_{\ell q}^{(3)}]_{3333}-[C_{\ell u}]_{3333}\right)\delta_{i3}\delta_{j3}\log \frac{\mu_{\rm EW}^2}{m_t^2},\\
[\delta g_{R\,ij}^{Ze}]_{\rm loop}=&\frac{3 m_t^2}{16 \pi^2} \left([\cC_{e q}]_{3333}-[\cC_{e u}]_{3333}\right)\delta_{i3}\delta_{j3}\log \frac{\mu_{\rm EW}^2}{m_t^2},\\
[\delta g_{L\, ij}^{Zu}]_{\rm loop}=&0,\\
[\delta g_{R\, ij}^{Zu}]_{\rm loop}=&0,\\
[\delta g_{L\,ij}^{Zd}]_{\rm loop}=&\frac{3 m_t^2}{16 \pi^2} \left(2[\cC_{qq}^{(1)}]_{3333}-2[\cC_{qq}^{(3)}]_{3333}
-[\cC_{q u}^{(1)}]_{3333}\right)\delta_{i3}\delta_{j3}\log \frac{\mu_{\rm EW}^2}{m_t^2}\nonumber\\
&+\frac{m_t^2}{4 \pi^2}[\cC_{qq}^{(3)}]_{3333}\delta_{i3}\delta_{j3}\left(-1+\log \frac{\mu_{\rm EW}^2}{m_t^2} \right)+\frac{m_t^2}{64\pi^2}[\cC_{Hu}]_{33}
\delta_{i3}\delta_{j3}\left( 1-2\log \frac{\mu_{\rm EW}^2}{m_t^2}\right),\label{eq:ZtobLbLLoop}\\
[\delta g_{R\, ij}^{Zd}]_{\rm loop}=&\,\frac{3 m_t^2}{16 \pi^2} \left([\cC^{(1)}_{qd}]_{3333}-[\cC^{(1)}_{ud}]_{3333}\right)\delta_{i3}\delta_{j3}\log \frac{\mu_{\rm EW}^2}{m_t^2}.
\end{align}
\noindent
Notice that $Z\to \bar b_L b_L$ in~\cref{eq:ZtobLbL} is only sensitive to the sum of $[\cC_{Hq}^{(1)}]$ and $[\cC_{Hq}^{(3)}]$. This sum contains finite pieces of the one-loop matching given in~\cref{CHq1FP,CHq3FP},
\begin{equation}
16\pi^2\Lambda_U^2\left([\cC_{Hq}^{(1)}]_{33}^{\rm{FP}}+[\cC_{Hq}^{(3)}]_{33}^{\rm{FP}}\right)=\frac{y_t^2}{ x_{G'}}-\frac{1}{8} \frac{\Lambda_U^2}{m_Q^2} y_t^2 |Y_+|^2.
\end{equation}
If we neglect the $y_t$ running, these finite pieces exactly cancel with the non-log terms of the one-loop correction to $Z\to \bar b_L b_L$ in~\cref{eq:ZtobLbLLoop}.
Moreover, when we insert the four-fermion operators at the tree level (see~\cref{tab:fourfermionoperators}) in the one-loop corrections to all vertex modifications relevant for the EW fit, $Z^{\prime}$ and $G^{\prime}$ contributions systematically cancel, leaving only the leptoquark contributions.

\noindent The variations of the $W$ couplings relevant for the EW fit (which excludes the $Wtb$ vertex), including the one-loop corrections in $y_t$ (diagram (b) of~\cref{fig:EW1loopdiagrams}) are
\begin{align}
\delta g^{W\ell}_{ij}=&\delta g_{L\,ij}^{Z\nu}-\delta g_{L\, ij}^{Ze}+\frac{3 m_t^2}{16 \pi^2}[\cC_{\ell q}^{(3)}]_{3333}\delta_{i3}\delta_{j3}, \label{eq:deltaWL}\\
\delta g^{Wq}_{ij}
=&\,\delta g_{L\,ik}^{Zu}V_{kj}-V_{ik}\delta g_{L\,kj}^{Zd}.
\end{align}
The $W$ mass modification, including the one-loop corrections in our model, is
\begin{align}
\delta m_W=&-\frac{v^2 g_L^2}{4(g_L^2-g_Y^2)}\cC_{HD}
-\frac{v^2 g_L g_Y}{g_L^2-g_Y^2}\cC_{HWB}
+\frac{v^2 g_Y^2}{4(g_L^2-g_Y^2)}\left([\cC_{\ell\ell}]_{1221}-2[\cC^{(3)}_{H\ell}]_{22}-2[\cC^{(3)}_{H\ell}]_{11}\right)\nonumber\\
&- \frac{g_L^2}{g_L^2-g_Y^2}\frac{3m_t^2}{8 \pi^2}[\cC_{Hu}]_{33} \log\frac{\mu_{\rm EW}^2}{m_t^2}.\label{eq:WmassExpr}
\end{align}

\subsection{Electroweak observable variations}
\noindent
To construct the likelihood for the EW fit, we use the observables $O$ of Tables 1 and 2 of~\cite{Breso-Pla:2021qoe}, that collect the measurements presented in \cite{ALEPH:2005ab,Janot:2019oyi,dEnterria:2020cgt,SLD:2000jop,
ParticleDataGroup:2020ssz,ALEPH:2013dgf,CDF:2005bdv,LHCb:2016zpq,
ATLAS:2016nqi,D0:1999bqi,ATLAS:2020xea}, and the SM predictions. Some sub-blocks of the observables are correlated. In \cref{tab:ZPoleCorrelations} and \cref{tab:WPoleCorrelations} we provide the correlations among these sub-blocks for the $Z$-pole observables~\cite{ALEPH:2005ab,Janot:2019oyi} and the $W$-pole observables~\cite{ALEPH:2013dgf} respectively.

\begin{table}[]
    \centering
    \renewcommand{\arraystretch}{1.4}
    \begin{tabular}{c |c c c c c c c c}
         & $\Gamma_Z$ & $\sigma_{\rm had}$ & $R_e$ & $R_{\mu}$ & $R_{\tau}$ & $A_{\rm FB}^{0,e}$ & $A_{\rm FB}^{0,\mu}$ & $A_{\rm FB}^{0,\tau}$ \\ \hline
   $\Gamma_Z$ &   1 & -0.3249 & -0.0110 & 0.0079 & 0.0059 & 0.0071 & 0.0020 & 0.0013 \\
    $\sigma_{\rm had}$ & &   1 & 0.1138 & 0.1391 & 0.0987 & 0.0015 & 0.0035 & 0.0018 \\
    $R_e$ & &   & 1 & 0.0694 & 0.0464 & -0.3704 & 0.0197 & 0.0132 \\
     $R_{\mu}$ & &   &  & 1 & 0.0696 & 0.0013 & 0.0121 & -0.0030 \\
      $R_{\tau}$ & &   &  &  & 1 & 0.0029 & 0.0012 & 0.0093 \\
     $A_{\rm FB}^{0,e}$ & &   &  &  &  & 1 & -0.0242 & -0.0202 \\
     $A_{\rm FB}^{0,\mu}$ & &   &  &  &  &  & 1 & 0.0464 \\
     $A_{\rm FB}^{0,\tau}$ & &   &  &  &  &  &  & 1 
    \end{tabular}
\\
\vspace{0.5cm}
    \begin{tabular}{c |c c c c}
         & $R_b$ & $R_c$ & $A^{\rm FB}_b$ & $A^{\rm FB}_{c}$ \\ \hline
   $R_b$ & 1 & -0.18 & -0.1 & 0.07 \\
   $R_c$ &           & 1 & 0.04 & -0.06 \\
   $A^{\rm FB}_b$      &     &  & 1 & 0.15 \\
   $A_{\rm FB}^{c}$    &       &  &  & 1 
    \end{tabular}
    \hspace{1cm}
    \begin{tabular}{c |c c c }
         & $A_e$ & $A_{\mu}$ & $A_{\tau}$  \\ \hline
      $A_e$ &       1 & 0.038 & 0.033 \\
      $A_{\mu}$ &       & 1 & 0.007 \\
      $A_{\tau}$ &           & & 1 
    \end{tabular}
\hspace{1cm}
    \begin{tabular}{c |c c }
          & $A_{b}$ & $A_{c}$  \\ \hline
      $A_b$ &       1 & 0.11 \\
      $A_c$ &       & 1 
  
\end{tabular}
    \caption{Correlations among the $Z$-pole EWPO~\cite{ALEPH:2005ab,Janot:2019oyi}.}
    \label{tab:ZPoleCorrelations}
\end{table}
\begin{table}[]
    \centering
    \renewcommand{\arraystretch}{1.4}
    \begin{tabular}{c |c c c }
         & Br$(W\to e\nu)$ & Br$(W\to \mu\nu)$ & Br$(W\to \tau\nu)$  \\ \hline
    Br$(W\to e\nu)$ &         1 & 0.136 & -0.201 \\
    Br$(W\to \mu\nu)$ &          & 1 & -0.122 \\
    Br$(W\to \tau\nu)$ &          &  & 1
    \end{tabular}
    \caption{Correlations among the $W$-pole EWPO~\cite{ALEPH:2013dgf}.}
    \label{tab:WPoleCorrelations}
\end{table}
We calculate the variation of the observables, $\Delta O =O_{\rm NP}-O_{\rm SM}$, at leading order in the EW boson vertex modifications and $\delta m_W$, as defined in~\cref{EWEffLag}, and neglecting the SM fermion masses. The expressions of the EW boson vertex modifications and $\delta m_W$ in terms of the SMEFT Wilson coefficients are given in~\cref{sec:SMEFTdelta}.
To write the variation of the observables, it is convenient first to define
\begin{align}
N_Z=&\frac{m_Z}{24 \pi}(g_L^2+g_Y^2),\\
[\Gamma_{Z}]_{\rm SM}=&\frac{N_Z}{12}\left( 63-120 s_W^2+160s_W^4 \right),\\
[\Gamma^{\rm had}_{Z}]_{\rm SM}=&\frac{N_Z}{12}\left(45-84 s_W^2+88s_W^4 \right),\\
\Delta \Gamma_Z^{\rm had} =&2N_Z\bigg[3\sum_{i=1}^2\left(g^{Zu}_L\delta g^{Zu}_{L\,ii}+g^{Zu}_R\delta g^{Zu}_{R\,ii} \right)+3\sum_{i=1}^3\left(g^{Zd}_L\delta g^{Zd}_{L\,ii}+g^{Zd}_R\delta g^{Zd}_{R\,ii} \right)
\bigg],\\
[\Gamma^{f}_{Z}]_{\rm SM}=&N_ZN_c^f\left[(g_L^{Zf})^2+
(g_R^{Zf})^2\right],\\
\Delta \Gamma^{f,i}_{Z}=&2N_ZN_c^f\left(g^{Zf}_L \delta g_{L\,ii}^{Zf}+g^{Zf}_R\delta g_{R\,ii}^{Zf}\right),\\
[A_{f}]_{\rm SM}=&\frac{(g^{Zf}_L)^2-(g^{Zf}_R)^2}{(g^{Zf}_L)^2+(g^{Zf}_R)^2},
\end{align}
where $N_c^f=1$ ($N_c^f=3$) for leptons (quarks) and $[O]_{\rm SM}$ represents the SM prediction of the observable $O$ at the tree level.
The variation of the $Z$-pole observables is then
\begin{align}
\Delta \Gamma_Z =&2N_Z \bigg[3\sum_{i=1}^2\left(g^{Zu}_L\delta g^{Zu}_{L\,ii}
+g^{Zu}_R\delta g^{Zu}_{R\,ii} \right)
+3\sum_{i=1}^3\left(g^{Zd}_L\delta g^{Zd}_{L\,ii}
+g^{Zd}_R\delta g^{Zd}_{R\,ii} \right)\nonumber\\
&~~~~~+\sum_{i=1}^3\left(g^{Ze}_L\delta g^{Ze}_{L\,ii}+ g^{Ze}_R \delta g^{Ze}_{R\,ii}+g^{Z\nu}_L\delta g^{Z\nu}_{L\,ii} \right)
\bigg],\\
\Delta \sigma_{\rm had}=&\frac{12\pi}{m_Z^2}\frac{[\Gamma_{Z}^{e}]_{\rm SM}[\Gamma_{Z}^{\rm had}]_{\rm SM}}{[\Gamma_{Z}]_{\rm SM}^2}\left(\frac{\Delta \Gamma_Z^{e,1}}{[\Gamma_{Z}^{e}]_{\rm SM}} +\frac{\Delta \Gamma_{Z}^{\rm had}}{[\Gamma_{Z}^{\rm had}]_{\rm SM}} -2\frac{\Delta\Gamma_{Z}}{[\Gamma_{Z}]_{\rm SM}} \right),\\
\Delta R_{e_i}=&\frac{\Delta \Gamma_{Z}^{\rm had}}{[\Gamma_{Z}^{e}]_{\rm SM}}-\frac{[\Gamma_{Z}^{\rm had}]_{\rm SM} \Delta \Gamma_Z^{e,i} }{[\Gamma_{Z}^{e}]_{\rm SM}^2},\\
\Delta R_{q_i}=&\frac{\Delta \Gamma_{Z}^{q,i}}{[\Gamma_{Z}^{\rm had}]_{\rm SM}}-\frac{[\Gamma_{Z}^{q}]_{\rm SM} \Delta \Gamma_Z^{\rm had} }{[\Gamma_{Z}^{\rm had}]_{\rm SM}^2},\\
\Delta A_{f_i}=&\frac{4g_{L}^{Zf}g_{R}^{Zf}}{\left((g_{L}^{Zf})^2+(g_{R}^{Zf})^2\right)^2}\left(
g_{R}^{Zf} \delta g_{L\,ii}^{Zf}-
g_{L}^{Zf} \delta g_{R\,ii}^{Zf} \right),\\
\Delta A_{\rm FB}^{0,e_{i}} =& \frac{3}{4}[A_e]_{\rm SM}(\Delta A_{e_1}+\Delta A_{e_i}),\\
\Delta A^{\rm FB}_{q_i} =& \frac{3}{4}\left([A_q]_{\rm SM}\Delta A_{e_1}+[A_e]_{\rm SM}\Delta A_{q_i}\right),\\
\Delta R_{uc}=&-\frac{[\Gamma_Z^{u}]_{\rm SM}}{ [\Gamma_Z^{\rm had}]^2_{\rm SM}}\Delta \Gamma_Z^{\rm had}+\frac{1}{2[\Gamma_Z^{\rm had}]_{\rm SM}}
\left(\Delta \Gamma_Z^{u,1} + \Delta \Gamma_Z^{u,2}\right).
\end{align}
The variation of the $W$-pole observables, working in the limit $V_{ij}=\delta_{ij}$, is
\begin{align}
\Delta\Gamma_{W}=&2N_W[m_W]_{\rm SM}\bigg(3\sum_{i=1}^2\delta g^{Wq}_{ii}+\sum_{i=1}^3\delta g^{W\ell}_{ii}
\bigg)+[\Gamma_W]_{\rm SM}\,\delta m_W,\\
\Delta {\rm Br}(W\to \ell_i\nu)=&
\frac{16}{81}\delta g_{ii}^{W\ell}-\frac{2}{81}\sum_{\substack{j=1\\j\neq i}}^3
\delta g_{jj}^{W\ell}-\frac{2}{27}\sum_{j=1,2} \delta g_{jj}^{Wq},\\
\Delta \frac{{\rm Br}(W\to \ell_i\nu)}{{\rm Br}(W\to \ell_j\nu)}=&2\delta g_{ii}^{W\ell}-2\delta g_{jj}^{W\ell},\\
\Delta R_{Wc}=&\frac{1}{2}\delta g_{22}^{Wq}-\frac{1}{2}\delta g_{11}^{Wq}.
\end{align}
where $N_W=g_L^2/48\pi$, $[m_W]_{\rm SM}=g_Lv/2$ and $[\Gamma_W]_{\rm SM}=9 N_W [m_W]_{\rm SM}$.
The effective number of neutrinos describes the $Z$ invisible width,
\begin{equation}
N_{\nu}^{\rm eff}=3\frac{\Gamma_{Z}^{\rm inv}}{\Gamma_{Z,{\rm SM}}^{\rm inv}}.
\end{equation}
It is not an observable of the EW fit itself, but can be calculated as a function of some of the EWPO~\cite{ALEPH:2005ab,Janot:2019oyi}. Its variation, as function of the $Z$-vertex modifications is
\begin{equation}
\Delta N_{\nu}^{\rm eff} = \frac{2}{g_L^{Z\nu}} \sum_{i=1}^3\delta g_{L\,ii}^{Z\nu}.
\end{equation}

\section{Expressions for other observables}\label{app:lowenergy}

\subsection{$b\to c\tau\bar\nu$ transitions}
\noindent
The Lagrangian relevant for $b\to c\tau\bar\nu$ transitions affecting the LFU ratios $R_{D^{(*)}}$, $R_{\Lambda_c}$ is
\begin{align}
   \cL_{b\to c\tau\Bar{\nu}}
  = - \frac{2}{v^2} V_{cb} & \bigg[
    \Big( 1 + \cC_{LL}^{c} \Big)
    (\bar c_L \gamma_\mu b_L) (\bar\tau_L \gamma^\mu \nu_L) - 2\,\cC_{LR}^{c} \,
    (\bar c_L b_R) (\bar\tau_R\,\nu_L) \bigg] \,,
    \label{eq:bctnuLag}
\end{align} 
so the LFU ratios read
\begin{align}
    \frac{R_D}{R_D^{\text{SM}}} &= \,\,  |1+\cC^c_{LL}|^2-3.00\, {\rm Re}\left[\left(1+\cC_{LL}^c\right) \cC^{c\,*}_{LR}\right]+4.12|\cC^c_{LR}|^2\,, \label{eq:RD} \\
    \frac{R_{D^*}}{R_{D^*}^{\text{SM}}} &=\,\,  |1+\cC^c_{LL}|^2-0.24\, {\rm Re}\left[\left(1+\cC_{LL}^c\right) \cC^{c\,*}_{LR}\right] +0.16| \cC^c_{LR}|^2\,, \\
    \frac{R_{\Lambda_c}}{R_{\Lambda_c}^{\rm SM}} &= \,\,
    |1+\cC^c_{LL}|^2-1.01\,{\rm Re}\left[\cC^{c}_{LR}+\cC_{LL}^c \cC^{c\,*}_{LR}\right]+1.34|\cC^c_{LR}|^2\,. \label{eq:RLambda}
\end{align}
The running can be neglected for $\cC_{LL}^c$, but not for $\cC_{LR}^c$. Including QCD running effects, 
\begin{equation}
\cC_{LR}^c(m_b)\approx 1.6 \, \cC_{LR}^c(\mu_{UV}).
\end{equation}
Defining the following Wilson coefficients
\begin{equation}
    \cC_{LL(LR)}^c = \cC_{LL(LR)}^{3333}\left[1+\frac{V_{cs}}{V_{cb}}\frac{\cC_{LL(LR)}^{2333}}{\cC_{LL(LR)}^{3333}}\left(1+ \frac{V_{ud}}{V_{cs}}\frac{\cC_{LL(LR)}^{1333}}{\cC_{LL(LR)}^{2333}}\right)\right]\,,
\end{equation}
the mapping to the Warsaw basis is
\begin{align}
      \cC_{LL}^{ij\alpha\beta} &= -\,v^2 \,[\cC_{\ell q}^{(3)}]_{\beta\alpha ij}\,,\\
      \cC_{LR}^{ij\alpha\beta} &=\,\frac{v^2}{4}\, [\cC_{\ell edq}]^*_{\alpha \beta ji}\,.
\end{align}

\subsection{$\tau$ LFU ratios}
\noindent
Following \cite{Allwicher:2021ndi}, the $\tau$ LFU ratios can be expressed in terms of SMEFT coefficients as
\begin{align}
    \nonumber
    \left(\frac{g_\tau}{g_{e,\mu}}\right)_{\ell,\pi,\mu} &\simeq 1 + v^2 {\rm Re} [\cC_{H\ell}^{(3)}]_{33}(\mu_{\rm EW}) + \frac{N_c m_t^2}{16\pi^2} {\rm Re} [\cC_{\ell q}^{(3)}]^{\rm tree}_{3333}\left(1+2 \log\frac{\mu_{\rm EW}^2}{m_t^2} \right) \\
    &= 1 + v^2 {\rm Re} [\cC_{H\ell}^{(3)}]_{33}(\mu_{\rm EW}) - \frac{N_c m_t^2 c_\chi^2}{32\pi^2\Lambda_U^2} 
    \left(1+2 \log\frac{\mu_{\rm EW}^2}{m_t^2} \right)\,,
\end{align}
where we have used the assumption that only $W$ couplings to taus are affected and substituted the tree-level matching for the $\cC_{\ell q}^{(3)}$ as in~\cref{tab:TLmatching}.

\bibliographystyle{JHEP}
\bibliography{references}

\end{document}